\documentclass[twocolumn,amsmath,amssymb,aps,prl,lettersize,superscriptaddress]{revtex4-1}

\usepackage{graphicx}
\usepackage{physics}
\usepackage{soul}
\usepackage{xcolor}
\usepackage{amsmath}
\usepackage{txfonts}
\usepackage{amssymb}
\usepackage{dcolumn}
\usepackage{bm}
\newcommand{\fig}[1]{{Fig.~\ref{#1}}}
\newcommand{\eq}[1]{{Eq.~(\ref{#1})}}

\usepackage{hyperref}
\hypersetup{colorlinks=true,citecolor=blue,linkcolor=blue,urlcolor=blue}

\begin{document}

\title{
Universal expansion of vortex clusters in a dissipative two-dimensional superfluid}

\author{Oliver R. Stockdale} \affiliation{ARC Centre of Excellence in Future Low-Energy Electronics Technologies, School of Mathematics and Physics, University of Queensland, St Lucia, QLD 4072, Australia}

\author{Matthew T. Reeves}\email{m.reeves@uq.edu.au}\affiliation{ARC Centre of Excellence in Future Low-Energy Electronics Technologies, School of Mathematics and Physics, University of Queensland, St Lucia, QLD 4072, Australia}

\author{Xiaoquan Yu}\email{xqyu@gscaep.ac.cn}
\affiliation{Graduate School of China Academy of Engineering Physics, Beijing 100193, China}
\affiliation{The Dodd-Walls Centre for Photonic and Quantum Technologies, Department of Physics, University of Otago, Dunedin 9016, New Zealand}

\author{Guillaume Gauthier}
\affiliation{ARC Centre of Excellence for Engineered Quantum Systems, School of Mathematics and Physics, University of Queensland, St Lucia, QLD 4072, Australia}

\author{Kwan Goddard-Lee}
\affiliation{ARC Centre of Excellence for Engineered Quantum Systems, School of Mathematics and Physics, University of Queensland, St Lucia, QLD 4072, Australia}

 \author{Warwick P. Bowen}
 \affiliation{ARC Centre of Excellence for Engineered Quantum Systems, School of Mathematics and Physics, University of Queensland, St Lucia, QLD 4072, Australia}

\author{Tyler W. Neely}
\affiliation{ARC Centre of Excellence for Engineered Quantum Systems, School of Mathematics and Physics, University of Queensland, St Lucia, QLD 4072, Australia}

\author{Matthew J. Davis}
\affiliation{ARC Centre of Excellence in Future Low-Energy Electronics Technologies, School of Mathematics and Physics, University of Queensland, St Lucia, QLD 4072, Australia}
\affiliation{ARC Centre of Excellence for Engineered Quantum Systems, School of Mathematics and Physics, University of Queensland, St Lucia, QLD 4072, Australia}

\date{\today}
\begin{abstract}
A large ensemble of quantum vortices in a superfluid may itself be treated as a novel kind of fluid that exhibits anomalous hydrodynamics. Here we consider the dynamics of vortex clusters with thermal friction, and present an analytic solution that uncovers a new universality class in the out-of-equilibrium dynamics of dissipative superfluids.  We find that the long-time dynamics of the vorticity distribution is an expanding Rankine vortex (i.e.~top-hat distribution) independent of initial conditions. This highlights a fundamentally different decay process to classical  fluids, where  the Rankine vortex is forbidden by viscous diffusion. Numerical simulations of large ensembles of point vortices confirm the universal expansion dynamics, and further reveal the emergence of a frustrated lattice structure marked by strong correlations. We present experimental results in a quasi-two-dimensional Bose-Einstein condensate that are in excellent agreement with the vortex fluid theory predictions, demonstrating that the signatures of vortex fluid theory can be observed with as few as $N\sim 11$ vortices. Our theoretical, numerical, and experimental results establish the validity of the vortex fluid theory for superfluid systems.

\end{abstract}

\maketitle

\textit{Introduction---}A defining feature of quantum fluids is that they exhibit quantized vortices. These stable topological defects have circulation that is quantized in units of $\Gamma~=~h/m$, where $h$ is Planck's constant and $m$ is the mass of a fluid particle. Despite this key difference from classical viscous fluids, many features of turbulence, i.e., the complex, collective behaviour of many vortices, are common to classical and quantum fluids. In three-dimensional quantum turbulence, which has been extensively studied in bulk superfluid helium~\cite{vinen2002,barenghi2014}, examples include the Kolmogorov energy cascade~\cite{proment2009,navon2016}, the dissipation anomaly~\cite{babuin2014}, and boundary layers~\cite{stagg2017}. More recently, experimental advances in two-dimensional (2D) ultracold atomic gases~\cite{hung2011,desbuquois2012,seo2017,mulkerin2017,ko2019} and superfluid optomechanical systems with thin-film helium~\cite{harris2016,sachkou2019} have renewed interest in turbulence and vortex dynamics in two dimensions, where markedly different behaviour to three dimensions is often observed. Here, 2D quantum fluid analogues of phenomena such as the von K{\'a}rm{\'a}n vortex street~\cite{sasaki2010,kwon2016}, and negative temperature vortex equilibria ~\cite{onsager1949,simula2014,yu2016,salman2016} have recently been demonstrated experimentally~\cite{gauthier2018,johnstone2018}.

One might expect the emergence of classical phenomena from quantum vortex dynamics to follow from Bohr's correspondence principle; provided many quantum vortices of the same sign are bundled together, collectively they should mimic classical vortex tubes. In two dimensions, however, recent theoretical work has shown that a dense system of chiral (i.e.\ same sign) quantum vortices at large scales may be treated as a kind of fluid in its own right.  In such a \emph{vortex fluid}, the dynamics are governed by a hydrodynamic equation that contains anomalous stress terms absent in the standard Euler equation~\cite{wiegmann2014}, allowing for phenomenon such as analogue edge states of the fractional quantum Hall effect~\cite{bogatskiy2019}. This theory was recently extended to describe dissipative effects~\cite{yu2017}, accounting for mutual friction due to interaction between the superfluid and a stationary thermal component present in experiments. However, exact solutions to the vortex fluid theory equations are still lacking, and an understanding of how the anomalous stresses affect large ensembles of quantum vortices remains unexplored. Furthermore, neither the conservative nor dissipative vortex fluid theory have been demonstrated experimentally.

Here we consider the dynamics of a 2D chiral vortex cluster within dissipative vortex fluid theory~\cite{yu2017}. Generally, solving for the out-of-equilibrium dynamics of many-body systems poses great challenges. We provide an analytical solution to this theory, demonstrating a new universality class in out-of-equilibrium dynamics of dissipative superfluids. We show that dense vortex clusters evolve into a Rankine vortex (i.e. top-hat distribution) at long times independent of the initial vorticity distribution. This behaviour is markedly different to the case of a classical viscous fluid, where not only is the Rankine vortex forbidden, but the expansion of a vortex is governed by ordinary viscous diffusion~\cite{lautrup2011}. 

To corroborate the theory, we simulate large collections of point vortices and provide strong evidence that any distribution of vorticity evolves into a Rankine vortex. Beyond the vortex fluid theory, we observe frustrated ordering of the vortices that become highly correlated at both short and long distances. Finally, we experimentally observe the emergence of a Rankine vortex in a $^{87}$Rb Bose-Einstein condensate with only $N\sim11$ vortices. Our findings establish a connection between the abstract concepts of the vortex fluid theory developed in Refs.~\cite{wiegmann2014,yu2017} and their physical realisations. Through our numerical and experimental results, we demonstrate a platform for further experiments investigating the vortex fluid theory.

\textit{Point vortex model---}We consider the motion of $N$ vortices in a homogeneous quasi-2D superfluid within the point vortex model~\cite{fetter1966, reeves2017}. Each vortex, at position $\mathbf{r}_i=(x_i,y_i)$, carries singly quantized circulation $\Gamma_i = \kappa_i h/m$ with $\kappa_i=\pm 1$ generating vorticity $\omega(\vb{r}) =\sum_i \Gamma_i\delta(\vb{r} - \vb{r}_i)$ and fluid velocity field  $\mathbf{u}(\vb{r})=2\pi\sum_i \Gamma_i\hat{z}\times(\vb{r}-\vb{r}_i)/|\vb{r}-\mathbf{r}_i|^2$. The incompressible kinetic energy of a 2D fluid can be expressed in terms of the relative vortex positions. In free space, the Hamiltonian is $H=-\rho_s\textstyle{\sum}_{i<j}\Gamma_i\Gamma_j\ln\qty(r_{ij}^2/L^2)/4\pi$, where $\rho_s$ is the 2D superfluid density, $r_{ij}=|\mathbf{r}_i-\mathbf{r}_j|$, and $L$ is an arbitrary length scale~\cite{aref2007}.  
Hamilton's equations give the velocity of vortex $i$ in terms of the other vortex positions as
\begin{align}
\mathbf{v}_{i} = \frac{1}{2\pi}
\sum_{j\neq i }\frac{\Gamma_j}{r_{ij}^2}
\begin{pmatrix}
-y_{ij}\\
x_{ij}\\
\end{pmatrix}, 
\label{eqnOfMotion}
\end{align}
where $x_{ij} = x_i-x_j$, $y_{ij} = y_i-y_j$, and $r_{ij}^2 = x_{ij}^2 + y_{ij}^2$.

For systems at finite temperature, interactions between the superfluid and thermal component result in the dissipation of energy~\cite{billam2015} proportional to the relative velocity of the two components. In two dimensions, the thermal component is typically stationary due to viscous clamping~\cite{harris2016}, or strong trap anisotropy~\cite{moon2015}, leading to the equation of motion
\begin{equation}
    \dot{\vb{r}}_i = \vb{v}_{i}  - \kappa_i  \gamma\, (\vu{z}\times  \vb{v}_{i}), 
    \label{eqn:dampedPVM}
\end{equation}
where the dimensionless mutual friction coefficient $\gamma$ (typically $\ll1 $) characterises the strength of the dissipation. Assuming superfluid density gradients are negligible, \eq{eqn:dampedPVM} can be rigorously derived from the damped Gross-Pitaevskii equation~\cite{tornkvist1997}, which quantitatively describes a weakly interacting BEC coupled to a uniform stationary thermal reservoir. In the context of superfluid helium, \eq{eqn:dampedPVM} and the mutual friction coefficient can also be rigorously derived from the interactions between a vortex and a thermal phonon bath~\cite{ambegaokar1978,ambegaokar1980,svistunov2015}.

\textit{Dissipative vortex fluid theory---}A system containing a large number of well-separated 2D quantum vortices can be viewed as a fluid in its own right, and its dynamics described by a set of hydrodynamic equations~\cite{wiegmann2014,yu2017}. For a chiral system where all vortices are of equal circulation ($\Gamma=h/m$), the collective dynamical variables are vortex density $\rho~\equiv~\sum_i \delta (\mathbf{r}-\mathbf{r}_i)$  and vortex fluid velocity field $\rho \mathbf{v}\equiv \sum_{i} \mathbf{v}_i \delta(\mathbf{r}-\mathbf{r}_i)$. Note that here $\omega=\Gamma \rho$. For a vortex fluid governed microscopically by \eq{eqn:dampedPVM}, the anomalous hydrodynamical equation for the vortex density is~\cite{yu2017}
\begin{align}
\mathcal{D}_t^v \rho =  -\gamma\left[\Gamma\rho^2+\frac{\Gamma}{8\pi}\nabla^2\rho-\mathbf{v}\times \nabla \rho -\frac{\Gamma}{8\pi}\frac{|\nabla\rho|^2}{\rho}\right] \label{yuHelmholtz},
\end{align}
where $\mathcal{D}_t^v \equiv \partial_t +\mathbf{v}\cdot \nabla$ is the material derivative, and $\gamma\mathbf{v}\times \nabla \rho$ describes transverse convection. The equation of motion for the vortex fluid velocity field $\mathbf{v}$ is complex and unimportant for our purposes here (see Supplementary Material~\cite{supp}).

In \eq{yuHelmholtz} the term $\propto \nabla^2 \rho$ describes \emph{uphill} diffusion of~$\rho$, which, in contrast to ordinary viscous diffusion, serves to steepen local vorticity gradients. The damping term $-\gamma\Gamma \rho^2$ strives to suppress regions of high density, and, together with the nonlinear term, balances the uphill diffusion to prevent a singular solution.

While \eq{yuHelmholtz} is intractable, by assuming uniform vortex density (due to the competition of terms), i.e. $\nabla \rho=0$, it simplifies to 
$
	\partial_t\rho = -\gamma\Gamma\rho^2,
$
which has the solution 
\begin{equation}
 \rho(t) = \left(\rho_0^{-1}+\gamma\Gamma t\right)^{-1},
\label{Rankine}
 \end{equation}
where $\rho_0 =\rho(0)$ is the initial density. This solution describes an expanding Rankine vortex, where the density distribution is uniform within the cluster, and zero outside~\footnote{Notice that a nontrivial vortex boundary layer may appear at the edge of the Rankine vortex~\cite{bogatskiy2019}, however this is beyond the scope of this work.}.
 
The Rankine vortex expansion is characterized by the mean radius
\begin{align}
\langle r(t)\rangle = \frac{2}{3}\sqrt{\frac{N}{\pi}}\left(\frac{1}{\rho_0}+\gamma\Gamma t\right)^{1/2}, \label{eqn:hydroRad}
\end{align}
which shows that the cluster exhibits diffusive-type growth. The canonical angular momentum of the fluid reads $L_f =-\int\dd\vb{r}\ r^2\omega/2 = -9\pi N\langle r(t)\rangle^2/8$, and  hence its dynamics is fully determined by \eq{eqn:hydroRad}.

The energy of the cluster, calculated from $H~=~\rho_s\int~\text{d}^2\mathbf{r}~|\vb{u}|^2/2$ with the velocity field of the Rankine vortex $u_{\phi}(r<r_c(t))=r \Gamma \rho/2$; $u_{\phi}(r>r_c(t))=\Gamma N/(2\pi r)$, evolves as 
\begin{align}
    H(t) = \frac{ \rho_s\Gamma^2N^2}{8\pi}\left(\frac{1}{2}-\ln\left[\frac{N}{\pi R^2}\qty(\frac{1}{\rho_0}+\gamma\Gamma t)\right]\right). \label{eqn:hydroEnergy}
\end{align}

We find that Eq.~\eqref{Rankine} is an asymptotic solution of Eq.~\eqref{yuHelmholtz} in the long time limit, independent of initial vortex distributions. The combination of the damping and negative viscosity terms in \eq{yuHelmholtz} suppresses density fluctuations in the cluster and yields the formation of a Rankine vortex that is an attractor of the dissipative dynamics. That is, an initially nonuniform density will evolve towards the universal Rankine scaling solution described by \eq{Rankine}. In classical viscous fluids, an axisymmetric vortex expands via diffusion, and  the late-time profile of an isolated line vortex instead tends to the Lamb-Oseen vortex. This distinct behaviour highlights the different dissipation mechanisms between finite temperature superfluids and classical viscous fluids. We find the Rankine vortex solution is robust against density fluctuations within perturbation analysis (see Supplementary Material~\cite{supp}) and further demonstrate its universality through numerical simulations.

\begin{figure}
    \centering
    \includegraphics[width=\columnwidth]{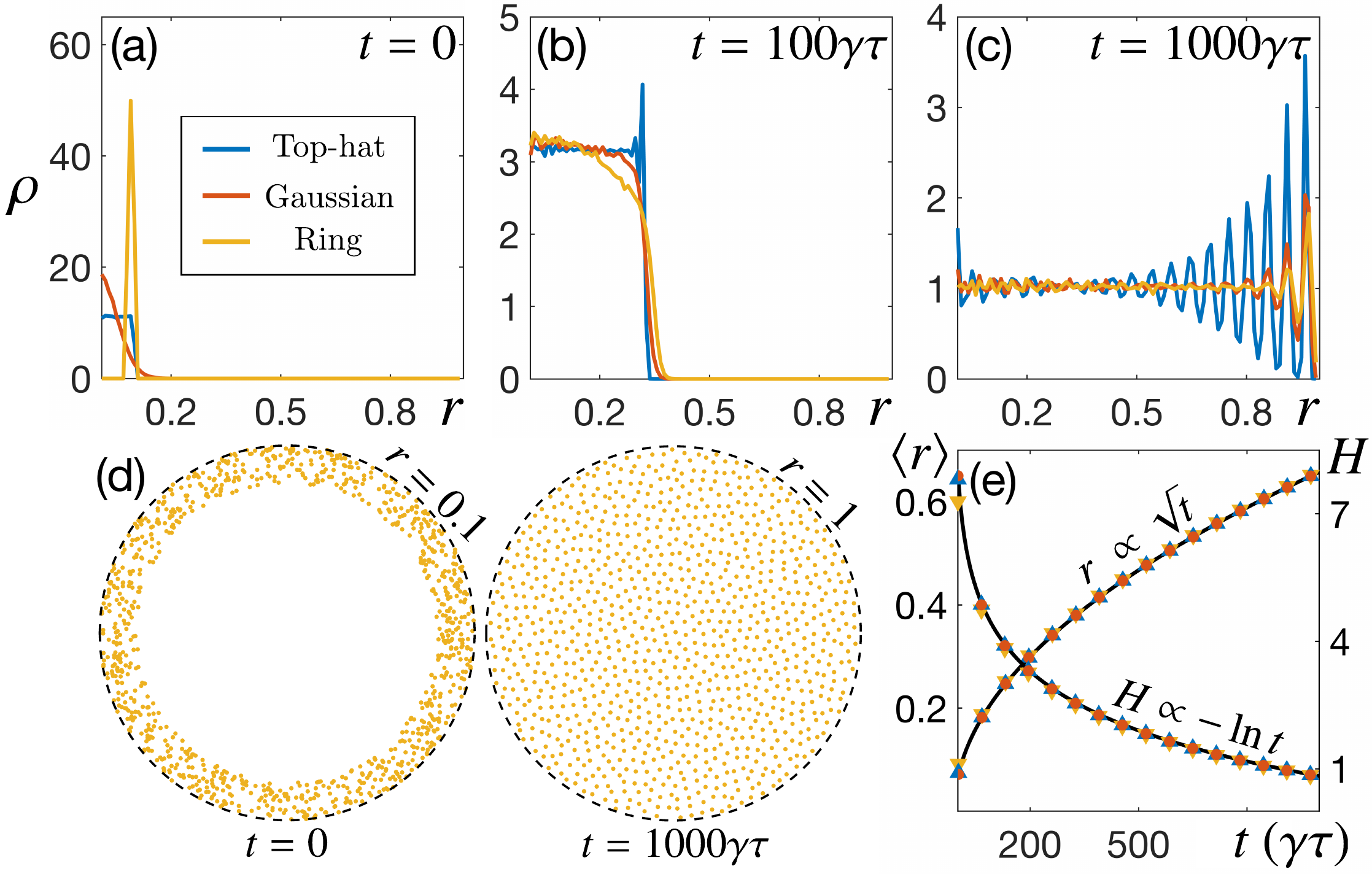}
    \caption{Ensemble radial vortex density for three different initial conditions at times (a) $t=0$, (b) $t=100\gamma\tau$, and (c) $t=1000\gamma\tau$.  (d) Example configurations of the ring ensemble at times in (a) and (c). Dashed lines around the cluster indicate the relative cluster size only. (e) Comparison between analytical vortex fluid theory (solid black curves) and numerical simulations.}
    \label{fig:densityPlots}
\end{figure}

\textit{Numerical results---}To demonstrate the universality numerically, in \fig{fig:densityPlots} we show the results of simulating the expansion of three ensembles of $N=1000$ vortices according to the point vortex model [\eq{eqn:dampedPVM}] with $\gamma=0.01$.  The initial conditions are drawn from top-hat, Gaussian, and ring distributions [Fig.~\ref{fig:densityPlots}(a)]. Without dissipation, the top-hat and Gaussian distributions are thermal equilibrium solutions for the vortex distribution~\cite{smith1990}. The ring is a highly non-equilibrium initial condition due to shear layer roll-up via the Kelvin-Helmholtz instability~\cite{smith1990,schecter1999}. 

The average radial density of each ensemble is shown for different times in Figs.~\ref{fig:densityPlots}(a--c), and exemplar configurations of the ring distribution are shown in Figs.~\ref{fig:densityPlots}(d).

The evolution of the cluster towards uniformity is seen in the radial density of the clusters at $t=100\gamma\tau$ and $t=1000\gamma\tau$ in Figs.~\ref{fig:densityPlots}(b,c), where we have chosen $\gamma\tau = 4.8\times10^{-2}~2\pi R^2/\Gamma$ as a convenient unit of time. The ring initial condition exhibits a slower timescale in evolving towards a top hat due to its highly non-uniform initial state. 
We find excellent agreement in the energy decay and mean radius growth between the simulation and the analytic solutions of the vortex fluid hydrodynamics as shown in \fig{fig:densityPlots}(e). The magnitude of $\gamma$ determines how quickly the cluster reaches the regime of universal dynamics, after which it is simply a scaling factor in time.

We have also simulated the expansion of several highly non-axisymmetric states and consistently found that each evolves to form a Rankine vortex~\cite{supp}. In the Supplemental Material we also provide comparisons with the conservative dynamics ($\gamma = 0$) for all scenarios~\cite{supp}.  The universal dynamics
is in contrast to the nonergodicity observed in conservative point vortex dynamics where large ensembles of point vortices can get trapped in an asymmetric, nonequilibrium stationary state~\cite{pakter2018}.

Once in the universal scaling regime, the coarse-grained vortex density (i.e., binned over scales larger than the inter-vortex distance) shows excellent agreement with the vortex fluid theory. At sufficiently late times, however, the structure begins to crystallize; additional short wavelength features emerge at length scales comparable to the typical inter-vortex distance, going beyond the vortex fluid theory predictions. Vortices at the cluster edge gradually organize into concentric circles, leading to a density peak at the boundary, with periodic oscillations decaying into the bulk [\fig{fig:densityPlots}(c), $t=1000\gamma\tau$]. We note these oscillations may be related to the predictions of Ref.~\cite{bogatskiy2019}, where it was found that the superfluid Rankine vortex supports an edge layer with a number of interesting properties, including a density overshoot and soliton solutions with quantized charge.
\begin{figure}
    \centering
    \includegraphics[width=\columnwidth]{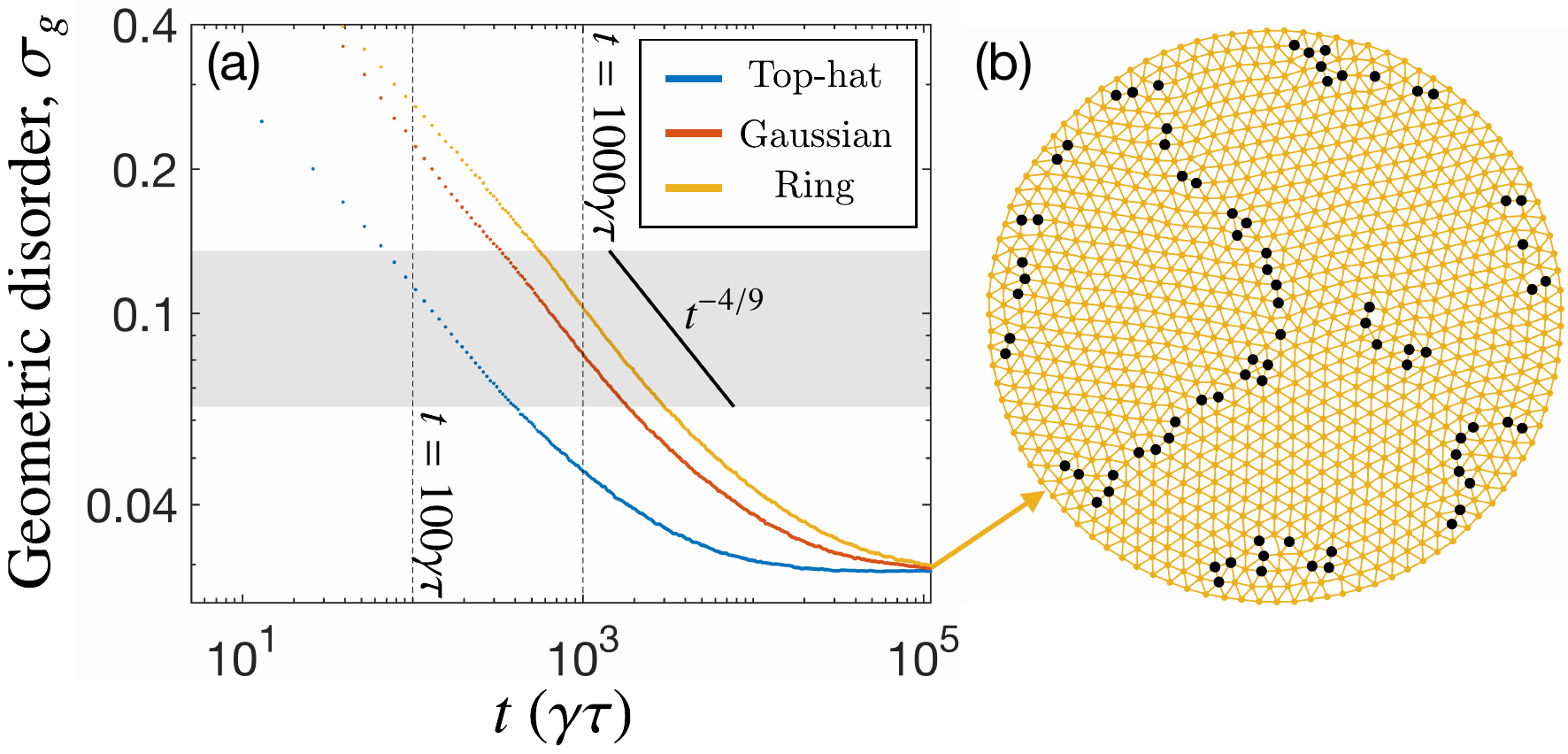}
    \caption{(a) Decay of the geometric disorder parameter, $\sigma_g$, averaged over $n=100$ runs. The shaded region represents a region where $\sigma_g\sim t^{-\alpha}$ with $\alpha\simeq 4/9$. (b) Exemplar cluster (ring initial distribution) at the limiting value of $\sigma_g$. Vortices are connected to their nearest neighbours to highlight structure, and black points represent vortices with five or seven nearest neighbours and indicate dislocations in the lattice.}
    \label{fig:correlation}
\end{figure}

In \fig{fig:correlation}(a) we quantify the dynamics of the crystallization by plotting the evolution of the geometric disorder parameter $\sigma_g = \sigma_{nn}/\mu_{nn}$ for the three examples of Fig.~\ref{fig:densityPlots}(a--c), where $\sigma_{nn}$ and $\mu_{nn}$ are the standard deviation and mean of the nearest neighbour distances of vortices in the cluster, respectively~\cite{rakonjac2016}. For a perfectly ordered Abrikosov lattice $\sigma_g=0$.  Interestingly, we find there is a period (indicated by the grey region) where all three ensembles evolve approximately as $\sigma_g\sim t^{-\alpha}$ with fitting parameter $\alpha\simeq 4/9$. Although we do not have an theoretical explanation for the observed scaling, we note that it was found to be independent of $\gamma$ for $\gamma\leq0.01$, suggesting it is a robust feature of the expansion. Furthermore, there is a late time plateau that persists for a wide range of $N$, suggesting $\lim_{t\rightarrow\infty}\sigma_g\neq0$ is not an artifact due to a particular choice of vortex number.  The tendency for the cluster to reach a steady-state where the disorder is non-zero suggests the outer concentric rings of vortices prevent the emerging Abrikosov lattice from spreading throughout the entire cluster. Clear signatures of an Abrikosov lattice can be seen in patches of the clusters [\fig{fig:correlation}(b)], however they are broken by dislocations that arise in the dynamics and persist. The cluster displays strong correlations at both short and long distances, owing to the local Abrikosov lattices and concentric outer rings respectively (see Supplementary Material~\cite{supp}). The frustrated structure observed is distinct from the familiar vortex lattices seen in rotating superfluid systems~\cite{haljan2001, abo2001,yarmchuk1979} as this system is out of equilibrium.

\begin{figure}
    \centering
    \includegraphics[width=\columnwidth]{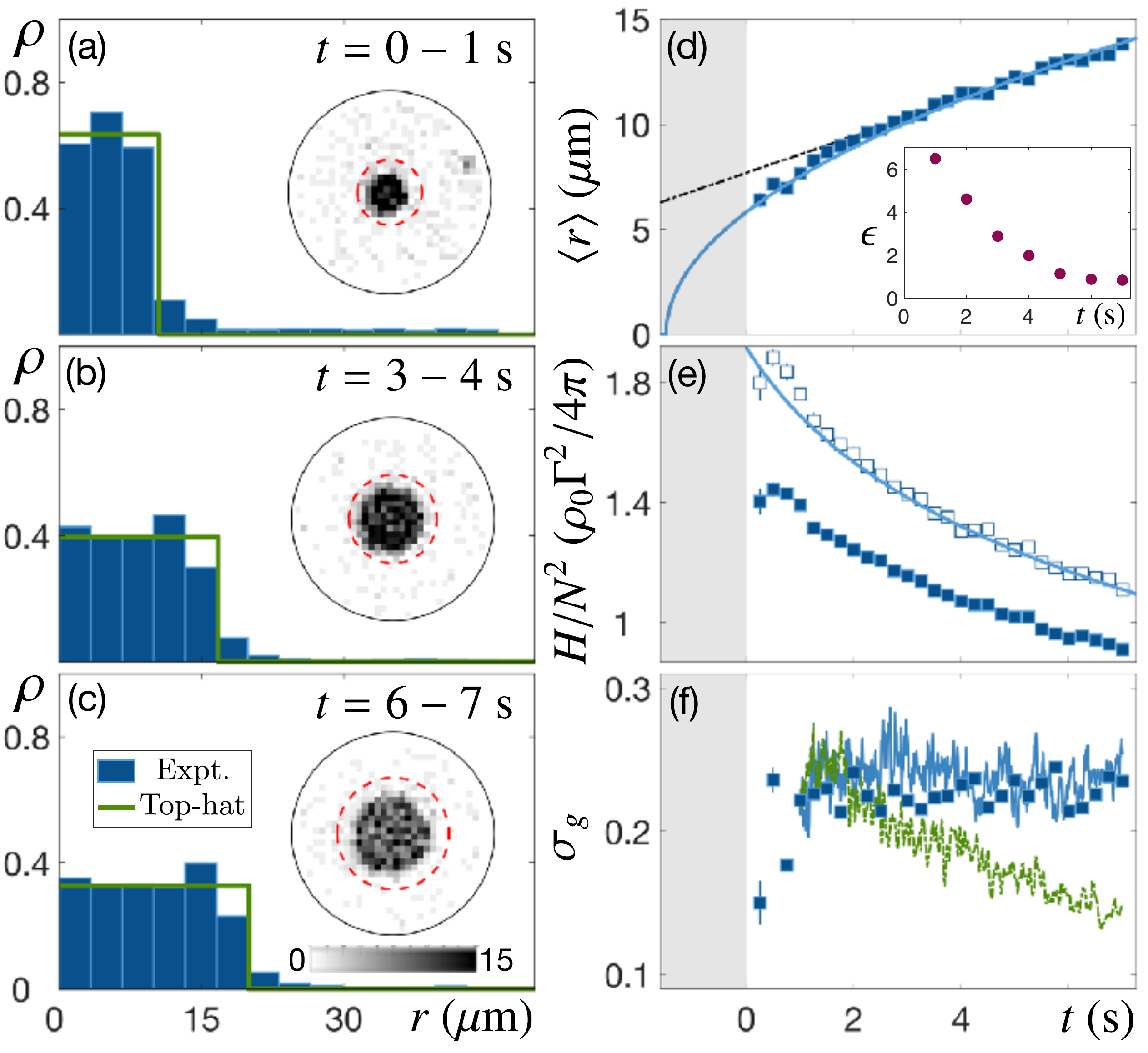}
    \caption{Expansion of a vortex cluster in a quasi-2D BEC. (a-c) Radial density of vortex clusters averaged over one second (period given in top right) along with top-hat distribution fit. Insets: corresponding 2D histograms of vortex positions. The red dashed line represents the cluster cutoff radius. (d) Inset: Percentage error between histograms and top-hat fit. Main: Comparison of vortex fluid theory (solid curve) and experimental measurements (points) of the average vortex radius $\langle r \rangle$. The dashed line is a linear fit to the late time data. The grey region is prior to the removal of the pinning potential. (e) Energy of experiment (solid points) and vortex fluid theory (solid curve). Experimental energy is also shown after being scaled (hollow points) by a constant factor (see text). (f) Observed geometric disorder $\sigma_g$ of the vortex cluster (points), compared with point vortex simulations with (solid blue) and without (dashed green) stray vortices used to calculate the dynamics.}
    \label{fig:experiment}
\end{figure}

\textit{Experiment---}Finally we compare the results of the vortex fluid theory to data from experiments on the expansion of vortex clusters in a quasi-2D BEC. A planar $^{87}$Rb BEC of $\sim2.2\times10^6$ atoms is confined to a ring-shaped trap of radius $R=50$~$\mu$m in the $x$--$y$ plane. The system is stirred, resulting in $N\sim 11$ same-sign vortices pinned to a central barrier, which is then slowly removed, leaving the eleven vortices clustered at the center of the nearly homogeneous superfluid (see~\cite{supp}). The system has a condensate fraction of $\sim80$\%, and the small thermal cloud leads to weakly dissipative vortex dynamics. Initially the vortices are within a radius of $\sim8~\mu$m, and are then free to expand in the presence of the dissipation. We observe the expansion for 6.75 seconds ($\sim4.1\times10^4~\gamma\tau$), destructively sampling $\sim40$ independent images at intervals of $250$~ms. A short 3~ms time-of-flight expands the vortex cores such that they are resolvable by our imaging, but it also means that not all eleven vortices can be resolved at early times (see~\cite{supp}).  We also observe occasional stray vortices away from the center of the superfluid that are created during the stirring protocol and are not centrally pinned (visible as a non-zero density outside the central cluster in \fig{fig:experiment}(a-c)).

In \fig{fig:experiment}(a-c) we plot the average radial density of the cluster for three different time intervals. In each case, we have fitted a top-hat distribution, where the radius is found from experimental measurements. We find that as the cluster expands, the density evolves towards a top-hat distribution [\fig{fig:experiment}(c)], which is further supported by the corresponding 2D histograms of vortex positions. Furthermore, we measure the error between the radial density and the top-hat fit and show it decreases and approaches zero~[Fig.~\ref{fig:experiment}(d) inset].

In \fig{fig:experiment}(d) we plot the mean radius of the central cluster as a function of time, and find excellent agreement with the $\langle r\rangle~\sim~\sqrt{t}$ prediction of the vortex fluid theory.  We fit a dissipation constant of $\gamma=3.2\times10^{-3}$ that is consistent with previous experiments~\cite{gauthier2018,moon2015,stagg2015}. Figure~\ref{fig:experiment}(e) shows the energy of the cluster. As the vortex fluid assumes $N \rightarrow \infty$, we observe a discrepancy in energy between experiment and theory due to only having $N\sim11$ vortices in the experiment. However, we find that applying a simple scaling factor $N^2/(N^2-2N)$ fits the theory well, and approaches unity as $N \rightarrow \infty$. Performing simulations of varying $N$, we found this scaling to yield good agreement with the fluid theory for all $N\geq10$~\cite{supp}. 

In \fig{fig:experiment}(f) we measure the disorder of the vortex cluster $\sigma_g$ experimentally, and compare it to simulations of the point vortex model [\eq{eqn:dampedPVM}] using the ensemble of initial vortex positions taken from the experiment at $t=1.0$~s, the earliest time that most vortex positions are resolved. If we only simulate the vortices in the initial cluster, we find the simulations predict a decrease in $\sigma_g$ with time, which is not consistent with the experimental data. However, if we include the additional stray vortices in the initial conditions, we find very good agreement, showing that the stray vortices significantly suppress the ordering of the central cluster. Without stray vortices, simulations show dissipation can provide sufficient ordering to study the crystallization dynamics in future experiments with larger vortex numbers.

\textit{Conclusions---}We have found that any dense vortex cluster in a finite temperature quantum fluid evolves to form a Rankine vortex confirming a new universality class in dissipative superfluids, as predicted by the dissipative vortex fluid theory. Beyond the universal expansion we find the emergence of frustrated lattice structure, which is approached through a power-law decay in the disorder. Finally, we have presented experimental observations of vortex cluster expansion in a quasi-2D BEC, and found that it is in good agreement with the vortex fluid theory despite being for a cluster of only eleven vortices. Our experimental results validate the vortex fluid theory, paving the way to better understand dissipative mechanisms in quantum fluids.

Whereas the Rankine vortex is forbidden in viscous classical fluids~\cite{lautrup2011}, our results suggest it may be highly relevant in finite temperature, dissipative superfluid systems. Our findings also suggest that recently predicted phenomena associated with the superfluid Rankine vortex, such as quantized edge solitons~\cite{bogatskiy2019}, may be within reach experimentally, provided larger vortex numbers can be achieved. Superfluids with a larger ratio between the system size and healing length, such as Fermi gases~\cite{ko2019} or thin-film superfluid helium~\cite{sachkou2019}, may be promising alternative platforms to test these predictions. Beyond the vortex fluid theory, emerging fractured lattice structure in the vortex clusters could have Kibble-Zurek type behaviour, as well demonstrating qualities reminiscent of the hexatic phase~\cite{nelson1979}. 

\begin{acknowledgments} We thank A.~S.~Bradley for useful discussions. This research was supported by the Australian Research Council Centre of Excellence in Future Low-Energy Electronics Technologies (project number CE170100039), the Australian Research Council Centre of Excellence for Engineered Quantum Systems (project number CE170100009), and funded by the Australian Government.  This work was also supported by the U.S. Army Research Office through grant number W911NF17-1-0310. X.Y. acknowledges the support from NSAF through grant number U1930403. O.R.S. acknowledges the support of an Australian Government Research and Training Program Scholarship.
\end{acknowledgments} 

\bibliography{references.bib}

\begin{thebibliography}{52}%
\makeatletter
\providecommand \@ifxundefined [1]{%
 \@ifx{#1\undefined}
}%
\providecommand \@ifnum [1]{%
 \ifnum #1\expandafter \@firstoftwo
 \else \expandafter \@secondoftwo
 \fi
}%
\providecommand \@ifx [1]{%
 \ifx #1\expandafter \@firstoftwo
 \else \expandafter \@secondoftwo
 \fi
}%
\providecommand \natexlab [1]{#1}%
\providecommand \enquote  [1]{``#1''}%
\providecommand \bibnamefont  [1]{#1}%
\providecommand \bibfnamefont [1]{#1}%
\providecommand \citenamefont [1]{#1}%
\providecommand \href@noop [0]{\@secondoftwo}%
\providecommand \href [0]{\begingroup \@sanitize@url \@href}%
\providecommand \@href[1]{\@@startlink{#1}\@@href}%
\providecommand \@@href[1]{\endgroup#1\@@endlink}%
\providecommand \@sanitize@url [0]{\catcode `\\12\catcode `\$12\catcode
  `\&12\catcode `\#12\catcode `\^12\catcode `\_12\catcode `\%12\relax}%
\providecommand \@@startlink[1]{}%
\providecommand \@@endlink[0]{}%
\providecommand \url  [0]{\begingroup\@sanitize@url \@url }%
\providecommand \@url [1]{\endgroup\@href {#1}{\urlprefix }}%
\providecommand \urlprefix  [0]{URL }%
\providecommand \Eprint [0]{\href }%
\providecommand \doibase [0]{http://dx.doi.org/}%
\providecommand \selectlanguage [0]{\@gobble}%
\providecommand \bibinfo  [0]{\@secondoftwo}%
\providecommand \bibfield  [0]{\@secondoftwo}%
\providecommand \translation [1]{[#1]}%
\providecommand \BibitemOpen [0]{}%
\providecommand \bibitemStop [0]{}%
\providecommand \bibitemNoStop [0]{.\EOS\space}%
\providecommand \EOS [0]{\spacefactor3000\relax}%
\providecommand \BibitemShut  [1]{\csname bibitem#1\endcsname}%
\let\auto@bib@innerbib\@empty
\bibitem [{\citenamefont {Vinen}\ and\ \citenamefont
  {Niemela}(2002)}]{vinen2002}%
  \BibitemOpen
  \bibfield  {author} {\bibinfo {author} {\bibfnamefont {W.~F.}\ \bibnamefont
  {Vinen}}\ and\ \bibinfo {author} {\bibfnamefont {J.~J.}\ \bibnamefont
  {Niemela}},\ }\href {\doibase 10.1023/A:1019695418590} {\bibfield  {journal}
  {\bibinfo  {journal} {Journal of Low Temperature Physics}\ }\textbf {\bibinfo
  {volume} {128}},\ \bibinfo {pages} {167} (\bibinfo {year}
  {2002})}\BibitemShut {NoStop}%
\bibitem [{\citenamefont {Barenghi}\ \emph {et~al.}(2014)\citenamefont
  {Barenghi}, \citenamefont {Skrbek},\ and\ \citenamefont
  {Sreenivasan}}]{barenghi2014}%
  \BibitemOpen
  \bibfield  {author} {\bibinfo {author} {\bibfnamefont {C.~F.}\ \bibnamefont
  {Barenghi}}, \bibinfo {author} {\bibfnamefont {L.}~\bibnamefont {Skrbek}}, \
  and\ \bibinfo {author} {\bibfnamefont {K.~R.}\ \bibnamefont {Sreenivasan}},\
  }\href {\doibase 10.1073/pnas.1400033111} {\bibfield  {journal} {\bibinfo
  {journal} {Proceedings of the National Academy of Sciences}\ ,\ \bibinfo
  {pages} {201400033}} (\bibinfo {year} {2014})}\BibitemShut {NoStop}%
\bibitem [{\citenamefont {Proment}\ \emph {et~al.}(2009)\citenamefont
  {Proment}, \citenamefont {Nazarenko},\ and\ \citenamefont
  {Onorato}}]{proment2009}%
  \BibitemOpen
  \bibfield  {author} {\bibinfo {author} {\bibfnamefont {D.}~\bibnamefont
  {Proment}}, \bibinfo {author} {\bibfnamefont {S.}~\bibnamefont {Nazarenko}},
  \ and\ \bibinfo {author} {\bibfnamefont {M.}~\bibnamefont {Onorato}},\ }\href
  {\doibase 10.1103/PhysRevA.80.051603} {\bibfield  {journal} {\bibinfo
  {journal} {Physical Review A}\ }\textbf {\bibinfo {volume} {80}},\ \bibinfo
  {pages} {051603(R)} (\bibinfo {year} {2009})}\BibitemShut {NoStop}%
\bibitem [{\citenamefont {Navon}\ \emph {et~al.}(2016)\citenamefont {Navon},
  \citenamefont {Gaunt}, \citenamefont {Smith},\ and\ \citenamefont
  {Hadzibabic}}]{navon2016}%
  \BibitemOpen
  \bibfield  {author} {\bibinfo {author} {\bibfnamefont {N.}~\bibnamefont
  {Navon}}, \bibinfo {author} {\bibfnamefont {A.~L.}\ \bibnamefont {Gaunt}},
  \bibinfo {author} {\bibfnamefont {R.~P.}\ \bibnamefont {Smith}}, \ and\
  \bibinfo {author} {\bibfnamefont {Z.}~\bibnamefont {Hadzibabic}},\ }\href
  {\doibase 10.1038/nature20114} {\bibfield  {journal} {\bibinfo  {journal}
  {Nature}\ }\textbf {\bibinfo {volume} {539}},\ \bibinfo {pages} {72}
  (\bibinfo {year} {2016})}\BibitemShut {NoStop}%
\bibitem [{\citenamefont {Babuin}\ \emph {et~al.}(2014)\citenamefont {Babuin},
  \citenamefont {Varga}, \citenamefont {Skrbek}, \citenamefont
  {L{\'e}v{\^e}que},\ and\ \citenamefont {Roche}}]{babuin2014}%
  \BibitemOpen
  \bibfield  {author} {\bibinfo {author} {\bibfnamefont {S.}~\bibnamefont
  {Babuin}}, \bibinfo {author} {\bibfnamefont {E.}~\bibnamefont {Varga}},
  \bibinfo {author} {\bibfnamefont {L.}~\bibnamefont {Skrbek}}, \bibinfo
  {author} {\bibfnamefont {E.}~\bibnamefont {L{\'e}v{\^e}que}}, \ and\ \bibinfo
  {author} {\bibfnamefont {P.-E.}\ \bibnamefont {Roche}},\ }\href {\doibase
  10.1209/0295-5075/106/24006} {\bibfield  {journal} {\bibinfo  {journal} {EPL
  (Europhysics Letters)}\ }\textbf {\bibinfo {volume} {106}},\ \bibinfo {pages}
  {24006} (\bibinfo {year} {2014})}\BibitemShut {NoStop}%
\bibitem [{\citenamefont {Stagg}\ \emph {et~al.}(2017)\citenamefont {Stagg},
  \citenamefont {Parker},\ and\ \citenamefont {Barenghi}}]{stagg2017}%
  \BibitemOpen
  \bibfield  {author} {\bibinfo {author} {\bibfnamefont {G.~W.}\ \bibnamefont
  {Stagg}}, \bibinfo {author} {\bibfnamefont {N.~G.}\ \bibnamefont {Parker}}, \
  and\ \bibinfo {author} {\bibfnamefont {C.~F.}\ \bibnamefont {Barenghi}},\
  }\href {\doibase 10.1103/PhysRevLett.118.135301} {\bibfield  {journal}
  {\bibinfo  {journal} {Physical Review Letters}\ }\textbf {\bibinfo {volume}
  {118}},\ \bibinfo {pages} {135301} (\bibinfo {year} {2017})}\BibitemShut
  {NoStop}%
\bibitem [{\citenamefont {Hung}\ \emph {et~al.}(2011)\citenamefont {Hung},
  \citenamefont {Zhang}, \citenamefont {Gemelke},\ and\ \citenamefont
  {Chin}}]{hung2011}%
  \BibitemOpen
  \bibfield  {author} {\bibinfo {author} {\bibfnamefont {C.-L.}\ \bibnamefont
  {Hung}}, \bibinfo {author} {\bibfnamefont {X.}~\bibnamefont {Zhang}},
  \bibinfo {author} {\bibfnamefont {N.}~\bibnamefont {Gemelke}}, \ and\
  \bibinfo {author} {\bibfnamefont {C.}~\bibnamefont {Chin}},\ }\href {\doibase
  10.1038/nature09722} {\bibfield  {journal} {\bibinfo  {journal} {Nature}\
  }\textbf {\bibinfo {volume} {470}},\ \bibinfo {pages} {236} (\bibinfo {year}
  {2011})}\BibitemShut {NoStop}%
\bibitem [{\citenamefont {Desbuquois}\ \emph {et~al.}(2012)\citenamefont
  {Desbuquois}, \citenamefont {Chomaz}, \citenamefont {Yefsah}, \citenamefont
  {L{\'e}onard}, \citenamefont {Beugnon}, \citenamefont {Weitenberg},\ and\
  \citenamefont {Dalibard}}]{desbuquois2012}%
  \BibitemOpen
  \bibfield  {author} {\bibinfo {author} {\bibfnamefont {R.}~\bibnamefont
  {Desbuquois}}, \bibinfo {author} {\bibfnamefont {L.}~\bibnamefont {Chomaz}},
  \bibinfo {author} {\bibfnamefont {T.}~\bibnamefont {Yefsah}}, \bibinfo
  {author} {\bibfnamefont {J.}~\bibnamefont {L{\'e}onard}}, \bibinfo {author}
  {\bibfnamefont {J.}~\bibnamefont {Beugnon}}, \bibinfo {author} {\bibfnamefont
  {C.}~\bibnamefont {Weitenberg}}, \ and\ \bibinfo {author} {\bibfnamefont
  {J.}~\bibnamefont {Dalibard}},\ }\href {\doibase 10.1038/nphys2378}
  {\bibfield  {journal} {\bibinfo  {journal} {Nature Physics}\ }\textbf
  {\bibinfo {volume} {8}},\ \bibinfo {pages} {645} (\bibinfo {year}
  {2012})}\BibitemShut {NoStop}%
\bibitem [{\citenamefont {Seo}\ \emph {et~al.}(2017)\citenamefont {Seo},
  \citenamefont {Ko}, \citenamefont {Kim},\ and\ \citenamefont
  {Shin}}]{seo2017}%
  \BibitemOpen
  \bibfield  {author} {\bibinfo {author} {\bibfnamefont {S.~W.}\ \bibnamefont
  {Seo}}, \bibinfo {author} {\bibfnamefont {B.}~\bibnamefont {Ko}}, \bibinfo
  {author} {\bibfnamefont {J.~H.}\ \bibnamefont {Kim}}, \ and\ \bibinfo
  {author} {\bibfnamefont {Y.-i.}\ \bibnamefont {Shin}},\ }\href {\doibase
  10.1038/s41598-017-04122-9} {\bibfield  {journal} {\bibinfo  {journal}
  {{Scientific Reports}}\ }\textbf {\bibinfo {volume} {7}},\ \bibinfo {pages}
  {4587} (\bibinfo {year} {2017})}\BibitemShut {NoStop}%
\bibitem [{\citenamefont {Mulkerin}\ \emph {et~al.}(2017)\citenamefont
  {Mulkerin}, \citenamefont {He}, \citenamefont {Dyke}, \citenamefont {Vale},
  \citenamefont {Liu},\ and\ \citenamefont {Hu}}]{mulkerin2017}%
  \BibitemOpen
  \bibfield  {author} {\bibinfo {author} {\bibfnamefont {B.~C.}\ \bibnamefont
  {Mulkerin}}, \bibinfo {author} {\bibfnamefont {L.}~\bibnamefont {He}},
  \bibinfo {author} {\bibfnamefont {P.}~\bibnamefont {Dyke}}, \bibinfo {author}
  {\bibfnamefont {C.~J.}\ \bibnamefont {Vale}}, \bibinfo {author}
  {\bibfnamefont {X.-J.}\ \bibnamefont {Liu}}, \ and\ \bibinfo {author}
  {\bibfnamefont {H.}~\bibnamefont {Hu}},\ }\href {\doibase
  10.1103/PhysRevA.96.053608} {\bibfield  {journal} {\bibinfo  {journal}
  {Physical Review A}\ }\textbf {\bibinfo {volume} {96}},\ \bibinfo {pages}
  {053608} (\bibinfo {year} {2017})}\BibitemShut {NoStop}%
\bibitem [{\citenamefont {Ko}\ \emph {et~al.}(2019)\citenamefont {Ko},
  \citenamefont {Park},\ and\ \citenamefont {Shin}}]{ko2019}%
  \BibitemOpen
  \bibfield  {author} {\bibinfo {author} {\bibfnamefont {B.}~\bibnamefont
  {Ko}}, \bibinfo {author} {\bibfnamefont {J.~W.}\ \bibnamefont {Park}}, \ and\
  \bibinfo {author} {\bibfnamefont {Y.-i.}\ \bibnamefont {Shin}},\ }\href
  {\doibase 10.1038/s41567-019-0650-1} {\bibfield  {journal} {\bibinfo
  {journal} {Nature Physics}\ }\textbf {\bibinfo {volume} {15}},\ \bibinfo
  {pages} {1227} (\bibinfo {year} {2019})}\BibitemShut {NoStop}%
\bibitem [{\citenamefont {Harris}\ \emph {et~al.}(2016)\citenamefont {Harris},
  \citenamefont {McAuslan}, \citenamefont {Sheridan}, \citenamefont {Sachkou},
  \citenamefont {Baker},\ and\ \citenamefont {Bowen}}]{harris2016}%
  \BibitemOpen
  \bibfield  {author} {\bibinfo {author} {\bibfnamefont {G.~I.}\ \bibnamefont
  {Harris}}, \bibinfo {author} {\bibfnamefont {D.~L.}\ \bibnamefont
  {McAuslan}}, \bibinfo {author} {\bibfnamefont {E.}~\bibnamefont {Sheridan}},
  \bibinfo {author} {\bibfnamefont {Y.~P.}\ \bibnamefont {Sachkou}}, \bibinfo
  {author} {\bibfnamefont {C.~G.}\ \bibnamefont {Baker}}, \ and\ \bibinfo
  {author} {\bibfnamefont {W.~P.}\ \bibnamefont {Bowen}},\ }\href {\doibase
  10.1038/nphys3714} {\bibfield  {journal} {\bibinfo  {journal} {Nature
  Physics}\ }\textbf {\bibinfo {volume} {12}},\ \bibinfo {pages} {788}
  (\bibinfo {year} {2016})}\BibitemShut {NoStop}%
\bibitem [{\citenamefont {Sachkou}\ \emph {et~al.}(2019)\citenamefont
  {Sachkou}, \citenamefont {Baker}, \citenamefont {Harris}, \citenamefont
  {Stockdale}, \citenamefont {Forstner}, \citenamefont {Reeves}, \citenamefont
  {He}, \citenamefont {McAuslan}, \citenamefont {Bradley}, \citenamefont
  {Davis},\ and\ \citenamefont {Bowen}}]{sachkou2019}%
  \BibitemOpen
  \bibfield  {author} {\bibinfo {author} {\bibfnamefont {Y.~P.}\ \bibnamefont
  {Sachkou}}, \bibinfo {author} {\bibfnamefont {C.~G.}\ \bibnamefont {Baker}},
  \bibinfo {author} {\bibfnamefont {G.~I.}\ \bibnamefont {Harris}}, \bibinfo
  {author} {\bibfnamefont {O.~R.}\ \bibnamefont {Stockdale}}, \bibinfo {author}
  {\bibfnamefont {S.}~\bibnamefont {Forstner}}, \bibinfo {author}
  {\bibfnamefont {M.~T.}\ \bibnamefont {Reeves}}, \bibinfo {author}
  {\bibfnamefont {X.}~\bibnamefont {He}}, \bibinfo {author} {\bibfnamefont
  {D.~L.}\ \bibnamefont {McAuslan}}, \bibinfo {author} {\bibfnamefont {A.~S.}\
  \bibnamefont {Bradley}}, \bibinfo {author} {\bibfnamefont {M.~J.}\
  \bibnamefont {Davis}}, \ and\ \bibinfo {author} {\bibfnamefont {W.~P.}\
  \bibnamefont {Bowen}},\ }\href {\doibase 10.1126/science.aaw9229} {\bibfield
  {journal} {\bibinfo  {journal} {Science}\ }\textbf {\bibinfo {volume}
  {366}},\ \bibinfo {pages} {1480} (\bibinfo {year} {2019})}\BibitemShut
  {NoStop}%
\bibitem [{\citenamefont {Sasaki}\ \emph {et~al.}(2010)\citenamefont {Sasaki},
  \citenamefont {Suzuki},\ and\ \citenamefont {Saito}}]{sasaki2010}%
  \BibitemOpen
  \bibfield  {author} {\bibinfo {author} {\bibfnamefont {K.}~\bibnamefont
  {Sasaki}}, \bibinfo {author} {\bibfnamefont {N.}~\bibnamefont {Suzuki}}, \
  and\ \bibinfo {author} {\bibfnamefont {H.}~\bibnamefont {Saito}},\ }\href
  {\doibase 10.1103/PhysRevLett.104.150404} {\bibfield  {journal} {\bibinfo
  {journal} {Physical Review Letters}\ }\textbf {\bibinfo {volume} {104}},\
  \bibinfo {pages} {150404} (\bibinfo {year} {2010})}\BibitemShut {NoStop}%
\bibitem [{\citenamefont {Kwon}\ \emph {et~al.}(2016)\citenamefont {Kwon},
  \citenamefont {Kim}, \citenamefont {Seo},\ and\ \citenamefont
  {Shin}}]{kwon2016}%
  \BibitemOpen
  \bibfield  {author} {\bibinfo {author} {\bibfnamefont {W.~J.}\ \bibnamefont
  {Kwon}}, \bibinfo {author} {\bibfnamefont {J.~H.}\ \bibnamefont {Kim}},
  \bibinfo {author} {\bibfnamefont {S.~W.}\ \bibnamefont {Seo}}, \ and\
  \bibinfo {author} {\bibfnamefont {Y.-i.}\ \bibnamefont {Shin}},\ }\href
  {\doibase 10.1103/PhysRevLett.117.245301} {\bibfield  {journal} {\bibinfo
  {journal} {Physical Review Letters}\ }\textbf {\bibinfo {volume} {117}},\
  \bibinfo {pages} {245301} (\bibinfo {year} {2016})}\BibitemShut {NoStop}%
\bibitem [{\citenamefont {Onsager}(1949)}]{onsager1949}%
  \BibitemOpen
  \bibfield  {author} {\bibinfo {author} {\bibfnamefont {L.}~\bibnamefont
  {Onsager}},\ }\href@noop {} {\bibfield  {journal} {\bibinfo  {journal} {Nuovo
  Cimento}\ }\textbf {\bibinfo {volume} {6}},\ \bibinfo {pages} {279} (\bibinfo
  {year} {1949})}\BibitemShut {NoStop}%
\bibitem [{\citenamefont {Simula}\ \emph {et~al.}(2014)\citenamefont {Simula},
  \citenamefont {Davis},\ and\ \citenamefont {Helmerson}}]{simula2014}%
  \BibitemOpen
  \bibfield  {author} {\bibinfo {author} {\bibfnamefont {T.}~\bibnamefont
  {Simula}}, \bibinfo {author} {\bibfnamefont {M.~J.}\ \bibnamefont {Davis}}, \
  and\ \bibinfo {author} {\bibfnamefont {K.}~\bibnamefont {Helmerson}},\ }\href
  {\doibase 10.1103/PhysRevLett.113.165302} {\bibfield  {journal} {\bibinfo
  {journal} {Physical Review Letters}\ }\textbf {\bibinfo {volume} {113}},\
  \bibinfo {pages} {165302} (\bibinfo {year} {2014})}\BibitemShut {NoStop}%
\bibitem [{\citenamefont {Yu}\ \emph {et~al.}(2016)\citenamefont {Yu},
  \citenamefont {Billam}, \citenamefont {Nian}, \citenamefont {Reeves},\ and\
  \citenamefont {Bradley}}]{yu2016}%
  \BibitemOpen
  \bibfield  {author} {\bibinfo {author} {\bibfnamefont {X.}~\bibnamefont
  {Yu}}, \bibinfo {author} {\bibfnamefont {T.~P.}\ \bibnamefont {Billam}},
  \bibinfo {author} {\bibfnamefont {J.}~\bibnamefont {Nian}}, \bibinfo {author}
  {\bibfnamefont {M.~T.}\ \bibnamefont {Reeves}}, \ and\ \bibinfo {author}
  {\bibfnamefont {A.~S.}\ \bibnamefont {Bradley}},\ }\href {\doibase
  10.1103/PhysRevA.94.023602} {\bibfield  {journal} {\bibinfo  {journal}
  {Physical Review A}\ }\textbf {\bibinfo {volume} {94}},\ \bibinfo {pages}
  {023602} (\bibinfo {year} {2016})}\BibitemShut {NoStop}%
\bibitem [{\citenamefont {Salman}\ and\ \citenamefont
  {Maestrini}(2016)}]{salman2016}%
  \BibitemOpen
  \bibfield  {author} {\bibinfo {author} {\bibfnamefont {H.}~\bibnamefont
  {Salman}}\ and\ \bibinfo {author} {\bibfnamefont {D.}~\bibnamefont
  {Maestrini}},\ }\href {\doibase 10.1103/PhysRevA.94.043642} {\bibfield
  {journal} {\bibinfo  {journal} {Physical Review A}\ }\textbf {\bibinfo
  {volume} {94}},\ \bibinfo {pages} {043642} (\bibinfo {year}
  {2016})}\BibitemShut {NoStop}%
\bibitem [{\citenamefont {Gauthier}\ \emph {et~al.}(2019)\citenamefont
  {Gauthier}, \citenamefont {Reeves}, \citenamefont {Yu}, \citenamefont
  {Bradley}, \citenamefont {Baker}, \citenamefont {Bell}, \citenamefont
  {Rubinsztein-Dunlop}, \citenamefont {Davis},\ and\ \citenamefont
  {Neely}}]{gauthier2018}%
  \BibitemOpen
  \bibfield  {author} {\bibinfo {author} {\bibfnamefont {G.}~\bibnamefont
  {Gauthier}}, \bibinfo {author} {\bibfnamefont {M.~T.}\ \bibnamefont
  {Reeves}}, \bibinfo {author} {\bibfnamefont {X.}~\bibnamefont {Yu}}, \bibinfo
  {author} {\bibfnamefont {A.~S.}\ \bibnamefont {Bradley}}, \bibinfo {author}
  {\bibfnamefont {M.~A.}\ \bibnamefont {Baker}}, \bibinfo {author}
  {\bibfnamefont {T.~A.}\ \bibnamefont {Bell}}, \bibinfo {author}
  {\bibfnamefont {H.}~\bibnamefont {Rubinsztein-Dunlop}}, \bibinfo {author}
  {\bibfnamefont {M.~J.}\ \bibnamefont {Davis}}, \ and\ \bibinfo {author}
  {\bibfnamefont {T.~W.}\ \bibnamefont {Neely}},\ }\href {\doibase
  10.1126/science.aat5718} {\bibfield  {journal} {\bibinfo  {journal}
  {Science}\ }\textbf {\bibinfo {volume} {364}},\ \bibinfo {pages} {1264}
  (\bibinfo {year} {2019})}\BibitemShut {NoStop}%
\bibitem [{\citenamefont {Johnstone}\ \emph {et~al.}(2019)\citenamefont
  {Johnstone}, \citenamefont {Groszek}, \citenamefont {Starkey}, \citenamefont
  {Billington}, \citenamefont {Simula},\ and\ \citenamefont
  {Helmerson}}]{johnstone2018}%
  \BibitemOpen
  \bibfield  {author} {\bibinfo {author} {\bibfnamefont {S.~P.}\ \bibnamefont
  {Johnstone}}, \bibinfo {author} {\bibfnamefont {A.~J.}\ \bibnamefont
  {Groszek}}, \bibinfo {author} {\bibfnamefont {P.~T.}\ \bibnamefont
  {Starkey}}, \bibinfo {author} {\bibfnamefont {C.~J.}\ \bibnamefont
  {Billington}}, \bibinfo {author} {\bibfnamefont {T.~P.}\ \bibnamefont
  {Simula}}, \ and\ \bibinfo {author} {\bibfnamefont {K.}~\bibnamefont
  {Helmerson}},\ }\href {\doibase 10.1126/science.aat5793} {\bibfield
  {journal} {\bibinfo  {journal} {Science}\ }\textbf {\bibinfo {volume}
  {364}},\ \bibinfo {pages} {1267} (\bibinfo {year} {2019})}\BibitemShut
  {NoStop}%
\bibitem [{\citenamefont {Wiegmann}\ and\ \citenamefont
  {Abanov}(2014)}]{wiegmann2014}%
  \BibitemOpen
  \bibfield  {author} {\bibinfo {author} {\bibfnamefont {P.}~\bibnamefont
  {Wiegmann}}\ and\ \bibinfo {author} {\bibfnamefont {A.~G.}\ \bibnamefont
  {Abanov}},\ }\href {\doibase 10.1103/PhysRevLett.113.034501} {\bibfield
  {journal} {\bibinfo  {journal} {Physical Review Letters}\ }\textbf {\bibinfo
  {volume} {113}},\ \bibinfo {pages} {034501} (\bibinfo {year}
  {2014})}\BibitemShut {NoStop}%
\bibitem [{\citenamefont {Bogatskiy}\ and\ \citenamefont
  {Wiegmann}(2019)}]{bogatskiy2019}%
  \BibitemOpen
  \bibfield  {author} {\bibinfo {author} {\bibfnamefont {A.}~\bibnamefont
  {Bogatskiy}}\ and\ \bibinfo {author} {\bibfnamefont {P.}~\bibnamefont
  {Wiegmann}},\ }\href {\doibase 10.1103/PhysRevLett.122.214505} {\bibfield
  {journal} {\bibinfo  {journal} {Physical Review Letters}\ }\textbf {\bibinfo
  {volume} {122}},\ \bibinfo {pages} {214505} (\bibinfo {year}
  {2019})}\BibitemShut {NoStop}%
\bibitem [{\citenamefont {Yu}\ and\ \citenamefont {Bradley}(2017)}]{yu2017}%
  \BibitemOpen
  \bibfield  {author} {\bibinfo {author} {\bibfnamefont {X.}~\bibnamefont
  {Yu}}\ and\ \bibinfo {author} {\bibfnamefont {A.~S.}\ \bibnamefont
  {Bradley}},\ }\href {\doibase 10.1103/PhysRevLett.119.185301} {\bibfield
  {journal} {\bibinfo  {journal} {Physical Review Letters}\ }\textbf {\bibinfo
  {volume} {119}},\ \bibinfo {pages} {185301} (\bibinfo {year}
  {2017})}\BibitemShut {NoStop}%
\bibitem [{\citenamefont {Lautrup}(2011)}]{lautrup2011}%
  \BibitemOpen
  \bibfield  {author} {\bibinfo {author} {\bibfnamefont {B.}~\bibnamefont
  {Lautrup}},\ }\href@noop {} {\emph {\bibinfo {title} {Physics of continuous
  matter: exotic and everyday phenomena in the macroscopic world}}}\ (\bibinfo
  {publisher} {CRC Press, Boca Raton},\ \bibinfo {year} {2011})\BibitemShut
  {NoStop}%
\bibitem [{\citenamefont {Fetter}(1966)}]{fetter1966}%
  \BibitemOpen
  \bibfield  {author} {\bibinfo {author} {\bibfnamefont {A.~L.}\ \bibnamefont
  {Fetter}},\ }\href {\doibase 10.1103/PhysRev.151.100} {\bibfield  {journal}
  {\bibinfo  {journal} {Physical Review}\ }\textbf {\bibinfo {volume} {151}},\
  \bibinfo {pages} {100} (\bibinfo {year} {1966})}\BibitemShut {NoStop}%
\bibitem [{\citenamefont {Reeves}\ \emph {et~al.}(2017)\citenamefont {Reeves},
  \citenamefont {Billam}, \citenamefont {Yu},\ and\ \citenamefont
  {Bradley}}]{reeves2017}%
  \BibitemOpen
  \bibfield  {author} {\bibinfo {author} {\bibfnamefont {M.~T.}\ \bibnamefont
  {Reeves}}, \bibinfo {author} {\bibfnamefont {T.~P.}\ \bibnamefont {Billam}},
  \bibinfo {author} {\bibfnamefont {X.}~\bibnamefont {Yu}}, \ and\ \bibinfo
  {author} {\bibfnamefont {A.~S.}\ \bibnamefont {Bradley}},\ }\href {\doibase
  10.1103/PhysRevLett.119.184502} {\bibfield  {journal} {\bibinfo  {journal}
  {Physical Review Letters}\ }\textbf {\bibinfo {volume} {119}},\ \bibinfo
  {pages} {184502} (\bibinfo {year} {2017})}\BibitemShut {NoStop}%
\bibitem [{\citenamefont {Aref}(2007)}]{aref2007}%
  \BibitemOpen
  \bibfield  {author} {\bibinfo {author} {\bibfnamefont {H.}~\bibnamefont
  {Aref}},\ }\href {\doibase 10.1063/1.2425103} {\bibfield  {journal} {\bibinfo
   {journal} {Journal of Mathematical Physics}\ }\textbf {\bibinfo {volume}
  {48}},\ \bibinfo {pages} {065401} (\bibinfo {year} {2007})}\BibitemShut
  {NoStop}%
\bibitem [{\citenamefont {Billam}\ \emph {et~al.}(2015)\citenamefont {Billam},
  \citenamefont {Reeves},\ and\ \citenamefont {Bradley}}]{billam2015}%
  \BibitemOpen
  \bibfield  {author} {\bibinfo {author} {\bibfnamefont {T.~P.}\ \bibnamefont
  {Billam}}, \bibinfo {author} {\bibfnamefont {M.~T.}\ \bibnamefont {Reeves}},
  \ and\ \bibinfo {author} {\bibfnamefont {A.~S.}\ \bibnamefont {Bradley}},\
  }\href {\doibase 10.1103/PhysRevA.91.023615} {\bibfield  {journal} {\bibinfo
  {journal} {{Physical Review A}}\ }\textbf {\bibinfo {volume} {91}},\ \bibinfo
  {pages} {023615} (\bibinfo {year} {2015})}\BibitemShut {NoStop}%
\bibitem [{\citenamefont {Moon}\ \emph {et~al.}(2015)\citenamefont {Moon},
  \citenamefont {Kwon}, \citenamefont {Lee},\ and\ \citenamefont
  {Shin}}]{moon2015}%
  \BibitemOpen
  \bibfield  {author} {\bibinfo {author} {\bibfnamefont {G.}~\bibnamefont
  {Moon}}, \bibinfo {author} {\bibfnamefont {W.~J.}\ \bibnamefont {Kwon}},
  \bibinfo {author} {\bibfnamefont {H.}~\bibnamefont {Lee}}, \ and\ \bibinfo
  {author} {\bibfnamefont {Y.-i.}\ \bibnamefont {Shin}},\ }\href {\doibase
  10.1103/PhysRevA.92.051601} {\bibfield  {journal} {\bibinfo  {journal}
  {Physical Review A}\ }\textbf {\bibinfo {volume} {92}},\ \bibinfo {pages}
  {051601(R)} (\bibinfo {year} {2015})}\BibitemShut {NoStop}%
\bibitem [{\citenamefont {T{\"o}rnkvist}\ and\ \citenamefont
  {Schr{\"o}der}(1997)}]{tornkvist1997}%
  \BibitemOpen
  \bibfield  {author} {\bibinfo {author} {\bibfnamefont {O.}~\bibnamefont
  {T{\"o}rnkvist}}\ and\ \bibinfo {author} {\bibfnamefont {E.}~\bibnamefont
  {Schr{\"o}der}},\ }\href {\doibase 10.1103/PhysRevLett.78.1908} {\bibfield
  {journal} {\bibinfo  {journal} {Physical Review Letters}\ }\textbf {\bibinfo
  {volume} {78}},\ \bibinfo {pages} {1908} (\bibinfo {year}
  {1997})}\BibitemShut {NoStop}%
\bibitem [{\citenamefont {Ambegaokar}\ \emph {et~al.}(1978)\citenamefont
  {Ambegaokar}, \citenamefont {Halperin}, \citenamefont {Nelson},\ and\
  \citenamefont {Siggia}}]{ambegaokar1978}%
  \BibitemOpen
  \bibfield  {author} {\bibinfo {author} {\bibfnamefont {V.}~\bibnamefont
  {Ambegaokar}}, \bibinfo {author} {\bibfnamefont {B.~I.}\ \bibnamefont
  {Halperin}}, \bibinfo {author} {\bibfnamefont {D.~R.}\ \bibnamefont
  {Nelson}}, \ and\ \bibinfo {author} {\bibfnamefont {E.~D.}\ \bibnamefont
  {Siggia}},\ }\href {\doibase 10.1103/PhysRevLett.40.783} {\bibfield
  {journal} {\bibinfo  {journal} {Physical Review Letters}\ }\textbf {\bibinfo
  {volume} {40}},\ \bibinfo {pages} {783} (\bibinfo {year} {1978})}\BibitemShut
  {NoStop}%
\bibitem [{\citenamefont {Ambegaokar}\ \emph {et~al.}(1980)\citenamefont
  {Ambegaokar}, \citenamefont {Halperin}, \citenamefont {Nelson},\ and\
  \citenamefont {Siggia}}]{ambegaokar1980}%
  \BibitemOpen
  \bibfield  {author} {\bibinfo {author} {\bibfnamefont {V.}~\bibnamefont
  {Ambegaokar}}, \bibinfo {author} {\bibfnamefont {B.~I.}\ \bibnamefont
  {Halperin}}, \bibinfo {author} {\bibfnamefont {D.~R.}\ \bibnamefont
  {Nelson}}, \ and\ \bibinfo {author} {\bibfnamefont {E.~D.}\ \bibnamefont
  {Siggia}},\ }\href {\doibase 10.1103/PhysRevB.21.1806} {\bibfield  {journal}
  {\bibinfo  {journal} {Physical Review B}\ }\textbf {\bibinfo {volume} {21}},\
  \bibinfo {pages} {1806} (\bibinfo {year} {1980})}\BibitemShut {NoStop}%
\bibitem [{\citenamefont {Svistunov}\ \emph {et~al.}(2015)\citenamefont
  {Svistunov}, \citenamefont {Babaev},\ and\ \citenamefont
  {Prokof'ev}}]{svistunov2015}%
  \BibitemOpen
  \bibfield  {author} {\bibinfo {author} {\bibfnamefont {B.~V.}\ \bibnamefont
  {Svistunov}}, \bibinfo {author} {\bibfnamefont {E.~S.}\ \bibnamefont
  {Babaev}}, \ and\ \bibinfo {author} {\bibfnamefont {N.~V.}\ \bibnamefont
  {Prokof'ev}},\ }\href@noop {} {\emph {\bibinfo {title} {Superfluid states of
  matter}}}\ (\bibinfo  {publisher} {CRC Press, Boca Raton},\ \bibinfo {year}
  {2015})\BibitemShut {NoStop}%
\bibitem [{sup()}]{supp}%
  \BibitemOpen
  \href@noop {} {}\bibinfo {howpublished} {{See Supplemental Material at [URL
  will be inserted by publisher] for full details of the vortex fluid theory,
  the perturbation analysis of the Rankine vortex solution, additional point
  vortex model simulations for zero dissipation and non-axisymmetric initial
  conditions, vortex cluster correlations, and further details of the
  experimental methods and results. The Supplemental Material also contains
  additional
  Refs.~\cite{skaugen2017,gauthier2016,eckel2014,wright2013,bradley1997,ellis1989,buhler2002}}}\BibitemShut
  {NoStop}%
\bibitem [{Note1()}]{Note1}%
  \BibitemOpen
  \bibinfo {note} {Notice that a nontrivial vortex boundary layer may appear at
  the edge of the Rankine vortex~\cite {bogatskiy2019}, however this is beyond
  the scope of this work.}\BibitemShut {Stop}%
\bibitem [{\citenamefont {Smith}\ and\ \citenamefont
  {O'Neil}(1990)}]{smith1990}%
  \BibitemOpen
  \bibfield  {author} {\bibinfo {author} {\bibfnamefont {R.~A.}\ \bibnamefont
  {Smith}}\ and\ \bibinfo {author} {\bibfnamefont {T.~M.}\ \bibnamefont
  {O'Neil}},\ }\href {\doibase 10.1063/1.859362} {\bibfield  {journal}
  {\bibinfo  {journal} {Physics of Fluids B: Plasma Physics}\ }\textbf
  {\bibinfo {volume} {2}},\ \bibinfo {pages} {2961} (\bibinfo {year}
  {1990})}\BibitemShut {NoStop}%
\bibitem [{\citenamefont {Schecter}\ \emph {et~al.}(1999)\citenamefont
  {Schecter}, \citenamefont {Dubin}, \citenamefont {Fine},\ and\ \citenamefont
  {Driscoll}}]{schecter1999}%
  \BibitemOpen
  \bibfield  {author} {\bibinfo {author} {\bibfnamefont {D.~A.}\ \bibnamefont
  {Schecter}}, \bibinfo {author} {\bibfnamefont {D.~H.~E.}\ \bibnamefont
  {Dubin}}, \bibinfo {author} {\bibfnamefont {K.~S.}\ \bibnamefont {Fine}}, \
  and\ \bibinfo {author} {\bibfnamefont {C.~F.}\ \bibnamefont {Driscoll}},\
  }\href {\doibase 10.1063/1.869961} {\bibfield  {journal} {\bibinfo  {journal}
  {Physics of Fluids}\ }\textbf {\bibinfo {volume} {11}},\ \bibinfo {pages}
  {905} (\bibinfo {year} {1999})}\BibitemShut {NoStop}%
\bibitem [{\citenamefont {Pakter}\ and\ \citenamefont
  {Levin}(2018)}]{pakter2018}%
  \BibitemOpen
  \bibfield  {author} {\bibinfo {author} {\bibfnamefont {R.}~\bibnamefont
  {Pakter}}\ and\ \bibinfo {author} {\bibfnamefont {Y.}~\bibnamefont {Levin}},\
  }\href {\doibase 10.1103/PhysRevLett.121.020602} {\bibfield  {journal}
  {\bibinfo  {journal} {Physical Review Letters}\ }\textbf {\bibinfo {volume}
  {121}},\ \bibinfo {pages} {020602} (\bibinfo {year} {2018})}\BibitemShut
  {NoStop}%
\bibitem [{\citenamefont {Rakonjac}\ \emph {et~al.}(2016)\citenamefont
  {Rakonjac}, \citenamefont {Marchant}, \citenamefont {Billam}, \citenamefont
  {Helm}, \citenamefont {Yu}, \citenamefont {Gardiner},\ and\ \citenamefont
  {Cornish}}]{rakonjac2016}%
  \BibitemOpen
  \bibfield  {author} {\bibinfo {author} {\bibfnamefont {A.}~\bibnamefont
  {Rakonjac}}, \bibinfo {author} {\bibfnamefont {A.~L.}\ \bibnamefont
  {Marchant}}, \bibinfo {author} {\bibfnamefont {T.~P.}\ \bibnamefont
  {Billam}}, \bibinfo {author} {\bibfnamefont {J.~L.}\ \bibnamefont {Helm}},
  \bibinfo {author} {\bibfnamefont {M.~M.~H.}\ \bibnamefont {Yu}}, \bibinfo
  {author} {\bibfnamefont {S.~A.}\ \bibnamefont {Gardiner}}, \ and\ \bibinfo
  {author} {\bibfnamefont {S.~L.}\ \bibnamefont {Cornish}},\ }\href {\doibase
  10.1103/PhysRevA.93.013607} {\bibfield  {journal} {\bibinfo  {journal}
  {Physical Review A}\ }\textbf {\bibinfo {volume} {93}},\ \bibinfo {pages}
  {013607} (\bibinfo {year} {2016})}\BibitemShut {NoStop}%
\bibitem [{\citenamefont {Haljan}\ \emph {et~al.}(2001)\citenamefont {Haljan},
  \citenamefont {Coddington}, \citenamefont {Engels},\ and\ \citenamefont
  {Cornell}}]{haljan2001}%
  \BibitemOpen
  \bibfield  {author} {\bibinfo {author} {\bibfnamefont {P.~C.}\ \bibnamefont
  {Haljan}}, \bibinfo {author} {\bibfnamefont {I.}~\bibnamefont {Coddington}},
  \bibinfo {author} {\bibfnamefont {P.}~\bibnamefont {Engels}}, \ and\ \bibinfo
  {author} {\bibfnamefont {E.~A.}\ \bibnamefont {Cornell}},\ }\href {\doibase
  10.1103/PhysRevLett.87.210403} {\bibfield  {journal} {\bibinfo  {journal}
  {Physical Review Letters}\ }\textbf {\bibinfo {volume} {87}},\ \bibinfo
  {pages} {210403} (\bibinfo {year} {2001})}\BibitemShut {NoStop}%
\bibitem [{\citenamefont {Abo-Shaeer}\ \emph {et~al.}(2001)\citenamefont
  {Abo-Shaeer}, \citenamefont {Raman}, \citenamefont {Vogels},\ and\
  \citenamefont {Ketterle}}]{abo2001}%
  \BibitemOpen
  \bibfield  {author} {\bibinfo {author} {\bibfnamefont {J.~R.}\ \bibnamefont
  {Abo-Shaeer}}, \bibinfo {author} {\bibfnamefont {C.}~\bibnamefont {Raman}},
  \bibinfo {author} {\bibfnamefont {J.~M.}\ \bibnamefont {Vogels}}, \ and\
  \bibinfo {author} {\bibfnamefont {W.}~\bibnamefont {Ketterle}},\ }\href
  {\doibase 10.1126/science.1060182} {\bibfield  {journal} {\bibinfo  {journal}
  {Science}\ }\textbf {\bibinfo {volume} {292}},\ \bibinfo {pages} {476}
  (\bibinfo {year} {2001})}\BibitemShut {NoStop}%
\bibitem [{\citenamefont {Yarmchuk}\ \emph {et~al.}(1979)\citenamefont
  {Yarmchuk}, \citenamefont {Gordon},\ and\ \citenamefont
  {Packard}}]{yarmchuk1979}%
  \BibitemOpen
  \bibfield  {author} {\bibinfo {author} {\bibfnamefont {E.~J.}\ \bibnamefont
  {Yarmchuk}}, \bibinfo {author} {\bibfnamefont {M.~J.~V.}\ \bibnamefont
  {Gordon}}, \ and\ \bibinfo {author} {\bibfnamefont {R.~E.}\ \bibnamefont
  {Packard}},\ }\href {\doibase 10.1103/PhysRevLett.43.214} {\bibfield
  {journal} {\bibinfo  {journal} {Physical Review Letters}\ }\textbf {\bibinfo
  {volume} {43}},\ \bibinfo {pages} {214} (\bibinfo {year} {1979})}\BibitemShut
  {NoStop}%
\bibitem [{\citenamefont {Stagg}\ \emph {et~al.}(2015)\citenamefont {Stagg},
  \citenamefont {Allen}, \citenamefont {Parker},\ and\ \citenamefont
  {Barenghi}}]{stagg2015}%
  \BibitemOpen
  \bibfield  {author} {\bibinfo {author} {\bibfnamefont {G.~W.}\ \bibnamefont
  {Stagg}}, \bibinfo {author} {\bibfnamefont {A.~J.}\ \bibnamefont {Allen}},
  \bibinfo {author} {\bibfnamefont {N.~G.}\ \bibnamefont {Parker}}, \ and\
  \bibinfo {author} {\bibfnamefont {C.~F.}\ \bibnamefont {Barenghi}},\ }\href
  {\doibase 10.1103/PhysRevA.91.013612} {\bibfield  {journal} {\bibinfo
  {journal} {Physical Review A}\ }\textbf {\bibinfo {volume} {91}},\ \bibinfo
  {pages} {013612} (\bibinfo {year} {2015})}\BibitemShut {NoStop}%
\bibitem [{\citenamefont {Nelson}\ and\ \citenamefont
  {Halperin}(1979)}]{nelson1979}%
  \BibitemOpen
  \bibfield  {author} {\bibinfo {author} {\bibfnamefont {D.~R.}\ \bibnamefont
  {Nelson}}\ and\ \bibinfo {author} {\bibfnamefont {B.~I.}\ \bibnamefont
  {Halperin}},\ }\href {\doibase 10.1103/PhysRevB.19.2457} {\bibfield
  {journal} {\bibinfo  {journal} {Physical Review B}\ }\textbf {\bibinfo
  {volume} {19}},\ \bibinfo {pages} {2457} (\bibinfo {year}
  {1979})}\BibitemShut {NoStop}%
\bibitem [{\citenamefont {Skaugen}\ and\ \citenamefont
  {Angheluta}(2017)}]{skaugen2017}%
  \BibitemOpen
  \bibfield  {author} {\bibinfo {author} {\bibfnamefont {A.}~\bibnamefont
  {Skaugen}}\ and\ \bibinfo {author} {\bibfnamefont {L.}~\bibnamefont
  {Angheluta}},\ }\href {\doibase 10.1103/PhysRevE.95.052144} {\bibfield
  {journal} {\bibinfo  {journal} {Physical Review E}\ }\textbf {\bibinfo
  {volume} {95}},\ \bibinfo {pages} {052144} (\bibinfo {year}
  {2017})}\BibitemShut {NoStop}%
\bibitem [{\citenamefont {Gauthier}\ \emph {et~al.}(2016)\citenamefont
  {Gauthier}, \citenamefont {Lenton}, \citenamefont {Parry}, \citenamefont
  {Baker}, \citenamefont {Davis}, \citenamefont {Rubinsztein-Dunlop},\ and\
  \citenamefont {Neely}}]{gauthier2016}%
  \BibitemOpen
  \bibfield  {author} {\bibinfo {author} {\bibfnamefont {G.}~\bibnamefont
  {Gauthier}}, \bibinfo {author} {\bibfnamefont {I.}~\bibnamefont {Lenton}},
  \bibinfo {author} {\bibfnamefont {N.~M.}\ \bibnamefont {Parry}}, \bibinfo
  {author} {\bibfnamefont {M.}~\bibnamefont {Baker}}, \bibinfo {author}
  {\bibfnamefont {M.~J.}\ \bibnamefont {Davis}}, \bibinfo {author}
  {\bibfnamefont {H.}~\bibnamefont {Rubinsztein-Dunlop}}, \ and\ \bibinfo
  {author} {\bibfnamefont {T.~W.}\ \bibnamefont {Neely}},\ }\href {\doibase
  10.1364/OPTICA.3.001136} {\bibfield  {journal} {\bibinfo  {journal} {Optica}\
  }\textbf {\bibinfo {volume} {3}},\ \bibinfo {pages} {1136} (\bibinfo {year}
  {2016})}\BibitemShut {NoStop}%
\bibitem [{\citenamefont {Eckel}\ \emph {et~al.}(2014)\citenamefont {Eckel},
  \citenamefont {Lee}, \citenamefont {Jendrzejewski}, \citenamefont {Murray},
  \citenamefont {Clark}, \citenamefont {Lobb}, \citenamefont {Phillips},
  \citenamefont {Edwards},\ and\ \citenamefont {Campbell}}]{eckel2014}%
  \BibitemOpen
  \bibfield  {author} {\bibinfo {author} {\bibfnamefont {S.}~\bibnamefont
  {Eckel}}, \bibinfo {author} {\bibfnamefont {J.~G.}\ \bibnamefont {Lee}},
  \bibinfo {author} {\bibfnamefont {F.}~\bibnamefont {Jendrzejewski}}, \bibinfo
  {author} {\bibfnamefont {N.}~\bibnamefont {Murray}}, \bibinfo {author}
  {\bibfnamefont {C.~W.}\ \bibnamefont {Clark}}, \bibinfo {author}
  {\bibfnamefont {C.~J.}\ \bibnamefont {Lobb}}, \bibinfo {author}
  {\bibfnamefont {W.~D.}\ \bibnamefont {Phillips}}, \bibinfo {author}
  {\bibfnamefont {M.}~\bibnamefont {Edwards}}, \ and\ \bibinfo {author}
  {\bibfnamefont {G.~K.}\ \bibnamefont {Campbell}},\ }\href {\doibase
  10.1038/nature12958} {\bibfield  {journal} {\bibinfo  {journal} {Nature}\
  }\textbf {\bibinfo {volume} {506}},\ \bibinfo {pages} {200} (\bibinfo {year}
  {2014})}\BibitemShut {NoStop}%
\bibitem [{\citenamefont {Wright}\ \emph {et~al.}(2013)\citenamefont {Wright},
  \citenamefont {Blakestad}, \citenamefont {Lobb}, \citenamefont {Phillips},\
  and\ \citenamefont {Campbell}}]{wright2013}%
  \BibitemOpen
  \bibfield  {author} {\bibinfo {author} {\bibfnamefont {K.~C.}\ \bibnamefont
  {Wright}}, \bibinfo {author} {\bibfnamefont {R.~B.}\ \bibnamefont
  {Blakestad}}, \bibinfo {author} {\bibfnamefont {C.~J.}\ \bibnamefont {Lobb}},
  \bibinfo {author} {\bibfnamefont {W.~D.}\ \bibnamefont {Phillips}}, \ and\
  \bibinfo {author} {\bibfnamefont {G.~K.}\ \bibnamefont {Campbell}},\ }\href
  {\doibase 10.1103/PhysRevLett.110.025302} {\bibfield  {journal} {\bibinfo
  {journal} {Physical Review Letters}\ }\textbf {\bibinfo {volume} {110}},\
  \bibinfo {pages} {025302} (\bibinfo {year} {2013})}\BibitemShut {NoStop}%
\bibitem [{\citenamefont {Bradley}\ \emph {et~al.}(1997)\citenamefont
  {Bradley}, \citenamefont {Sackett},\ and\ \citenamefont
  {Hulet}}]{bradley1997}%
  \BibitemOpen
  \bibfield  {author} {\bibinfo {author} {\bibfnamefont {C.~C.}\ \bibnamefont
  {Bradley}}, \bibinfo {author} {\bibfnamefont {C.~A.}\ \bibnamefont
  {Sackett}}, \ and\ \bibinfo {author} {\bibfnamefont {R.~G.}\ \bibnamefont
  {Hulet}},\ }\href {\doibase 10.1103/PhysRevLett.78.985} {\bibfield  {journal}
  {\bibinfo  {journal} {Physical Review Letters}\ }\textbf {\bibinfo {volume}
  {78}},\ \bibinfo {pages} {985} (\bibinfo {year} {1997})}\BibitemShut
  {NoStop}%
\bibitem [{\citenamefont {Ellis}\ and\ \citenamefont {Luo}(1989)}]{ellis1989}%
  \BibitemOpen
  \bibfield  {author} {\bibinfo {author} {\bibfnamefont {F.~M.}\ \bibnamefont
  {Ellis}}\ and\ \bibinfo {author} {\bibfnamefont {H.}~\bibnamefont {Luo}},\
  }\href {\doibase 10.1103/PhysRevB.39.2703} {\bibfield  {journal} {\bibinfo
  {journal} {Physical Review B}\ }\textbf {\bibinfo {volume} {39}},\ \bibinfo
  {pages} {2703} (\bibinfo {year} {1989})}\BibitemShut {NoStop}%
\bibitem [{\citenamefont {B{\"u}hler}(2002)}]{buhler2002}%
  \BibitemOpen
  \bibfield  {author} {\bibinfo {author} {\bibfnamefont {O.}~\bibnamefont
  {B{\"u}hler}},\ }\href {\doibase 10.1063/1.1483305} {\bibfield  {journal}
  {\bibinfo  {journal} {Physics of Fluids}\ }\textbf {\bibinfo {volume} {14}},\
  \bibinfo {pages} {2139} (\bibinfo {year} {2002})}\BibitemShut {NoStop}%
\end{thebibliography}%


\begin{thebibliography}{11}%
\makeatletter
\providecommand \@ifxundefined [1]{%
 \@ifx{#1\undefined}
}%
\providecommand \@ifnum [1]{%
 \ifnum #1\expandafter \@firstoftwo
 \else \expandafter \@secondoftwo
 \fi
}%
\providecommand \@ifx [1]{%
 \ifx #1\expandafter \@firstoftwo
 \else \expandafter \@secondoftwo
 \fi
}%
\providecommand \natexlab [1]{#1}%
\providecommand \enquote  [1]{``#1''}%
\providecommand \bibnamefont  [1]{#1}%
\providecommand \bibfnamefont [1]{#1}%
\providecommand \citenamefont [1]{#1}%
\providecommand \href@noop [0]{\@secondoftwo}%
\providecommand \href [0]{\begingroup \@sanitize@url \@href}%
\providecommand \@href[1]{\@@startlink{#1}\@@href}%
\providecommand \@@href[1]{\endgroup#1\@@endlink}%
\providecommand \@sanitize@url [0]{\catcode `\\12\catcode `\$12\catcode
  `\&12\catcode `\#12\catcode `\^12\catcode `\_12\catcode `\%12\relax}%
\providecommand \@@startlink[1]{}%
\providecommand \@@endlink[0]{}%
\providecommand \url  [0]{\begingroup\@sanitize@url \@url }%
\providecommand \@url [1]{\endgroup\@href {#1}{\urlprefix }}%
\providecommand \urlprefix  [0]{URL }%
\providecommand \Eprint [0]{\href }%
\providecommand \doibase [0]{http://dx.doi.org/}%
\providecommand \selectlanguage [0]{\@gobble}%
\providecommand \bibinfo  [0]{\@secondoftwo}%
\providecommand \bibfield  [0]{\@secondoftwo}%
\providecommand \translation [1]{[#1]}%
\providecommand \BibitemOpen [0]{}%
\providecommand \bibitemStop [0]{}%
\providecommand \bibitemNoStop [0]{.\EOS\space}%
\providecommand \EOS [0]{\spacefactor3000\relax}%
\providecommand \BibitemShut  [1]{\csname bibitem#1\endcsname}%
\let\auto@bib@innerbib\@empty
\bibitem [{\citenamefont {Yu}\ and\ \citenamefont {Bradley}(2017)}]{yu2017}%
  \BibitemOpen
  \bibfield  {author} {\bibinfo {author} {\bibfnamefont {X.}~\bibnamefont
  {Yu}}\ and\ \bibinfo {author} {\bibfnamefont {A.~S.}\ \bibnamefont
  {Bradley}},\ }\href {\doibase 10.1103/PhysRevLett.119.185301} {\bibfield
  {journal} {\bibinfo  {journal} {Physical Review Letters}\ }\textbf {\bibinfo
  {volume} {119}},\ \bibinfo {pages} {185301} (\bibinfo {year}
  {2017})}\BibitemShut {NoStop}%
\bibitem [{\citenamefont {Skaugen}\ and\ \citenamefont
  {Angheluta}(2017)}]{skaugen2017}%
  \BibitemOpen
  \bibfield  {author} {\bibinfo {author} {\bibfnamefont {A.}~\bibnamefont
  {Skaugen}}\ and\ \bibinfo {author} {\bibfnamefont {L.}~\bibnamefont
  {Angheluta}},\ }\href {\doibase 10.1103/PhysRevE.95.052144} {\bibfield
  {journal} {\bibinfo  {journal} {Physical Review E}\ }\textbf {\bibinfo
  {volume} {95}},\ \bibinfo {pages} {052144} (\bibinfo {year}
  {2017})}\BibitemShut {NoStop}%
\bibitem [{\citenamefont {Gauthier}\ \emph {et~al.}(2016)\citenamefont
  {Gauthier}, \citenamefont {Lenton}, \citenamefont {Parry}, \citenamefont
  {Baker}, \citenamefont {Davis}, \citenamefont {Rubinsztein-Dunlop},\ and\
  \citenamefont {Neely}}]{gauthier2016}%
  \BibitemOpen
  \bibfield  {author} {\bibinfo {author} {\bibfnamefont {G.}~\bibnamefont
  {Gauthier}}, \bibinfo {author} {\bibfnamefont {I.}~\bibnamefont {Lenton}},
  \bibinfo {author} {\bibfnamefont {N.~M.}\ \bibnamefont {Parry}}, \bibinfo
  {author} {\bibfnamefont {M.}~\bibnamefont {Baker}}, \bibinfo {author}
  {\bibfnamefont {M.~J.}\ \bibnamefont {Davis}}, \bibinfo {author}
  {\bibfnamefont {H.}~\bibnamefont {Rubinsztein-Dunlop}}, \ and\ \bibinfo
  {author} {\bibfnamefont {T.~W.}\ \bibnamefont {Neely}},\ }\href {\doibase
  10.1364/OPTICA.3.001136} {\bibfield  {journal} {\bibinfo  {journal} {Optica}\
  }\textbf {\bibinfo {volume} {3}},\ \bibinfo {pages} {1136} (\bibinfo {year}
  {2016})}\BibitemShut {NoStop}%
\bibitem [{\citenamefont {Gauthier}\ \emph {et~al.}(2019)\citenamefont
  {Gauthier}, \citenamefont {Reeves}, \citenamefont {Yu}, \citenamefont
  {Bradley}, \citenamefont {Baker}, \citenamefont {Bell}, \citenamefont
  {Rubinsztein-Dunlop}, \citenamefont {Davis},\ and\ \citenamefont
  {Neely}}]{gauthier2018}%
  \BibitemOpen
  \bibfield  {author} {\bibinfo {author} {\bibfnamefont {G.}~\bibnamefont
  {Gauthier}}, \bibinfo {author} {\bibfnamefont {M.~T.}\ \bibnamefont
  {Reeves}}, \bibinfo {author} {\bibfnamefont {X.}~\bibnamefont {Yu}}, \bibinfo
  {author} {\bibfnamefont {A.~S.}\ \bibnamefont {Bradley}}, \bibinfo {author}
  {\bibfnamefont {M.~A.}\ \bibnamefont {Baker}}, \bibinfo {author}
  {\bibfnamefont {T.~A.}\ \bibnamefont {Bell}}, \bibinfo {author}
  {\bibfnamefont {H.}~\bibnamefont {Rubinsztein-Dunlop}}, \bibinfo {author}
  {\bibfnamefont {M.~J.}\ \bibnamefont {Davis}}, \ and\ \bibinfo {author}
  {\bibfnamefont {T.~W.}\ \bibnamefont {Neely}},\ }\href {\doibase
  10.1126/science.aat5718} {\bibfield  {journal} {\bibinfo  {journal}
  {Science}\ }\textbf {\bibinfo {volume} {364}},\ \bibinfo {pages} {1264}
  (\bibinfo {year} {2019})}\BibitemShut {NoStop}%
\bibitem [{\citenamefont {Eckel}\ \emph {et~al.}(2014)\citenamefont {Eckel},
  \citenamefont {Lee}, \citenamefont {Jendrzejewski}, \citenamefont {Murray},
  \citenamefont {Clark}, \citenamefont {Lobb}, \citenamefont {Phillips},
  \citenamefont {Edwards},\ and\ \citenamefont {Campbell}}]{eckel2014}%
  \BibitemOpen
  \bibfield  {author} {\bibinfo {author} {\bibfnamefont {S.}~\bibnamefont
  {Eckel}}, \bibinfo {author} {\bibfnamefont {J.~G.}\ \bibnamefont {Lee}},
  \bibinfo {author} {\bibfnamefont {F.}~\bibnamefont {Jendrzejewski}}, \bibinfo
  {author} {\bibfnamefont {N.}~\bibnamefont {Murray}}, \bibinfo {author}
  {\bibfnamefont {C.~W.}\ \bibnamefont {Clark}}, \bibinfo {author}
  {\bibfnamefont {C.~J.}\ \bibnamefont {Lobb}}, \bibinfo {author}
  {\bibfnamefont {W.~D.}\ \bibnamefont {Phillips}}, \bibinfo {author}
  {\bibfnamefont {M.}~\bibnamefont {Edwards}}, \ and\ \bibinfo {author}
  {\bibfnamefont {G.~K.}\ \bibnamefont {Campbell}},\ }\href {\doibase
  10.1038/nature12958} {\bibfield  {journal} {\bibinfo  {journal} {Nature}\
  }\textbf {\bibinfo {volume} {506}},\ \bibinfo {pages} {200} (\bibinfo {year}
  {2014})}\BibitemShut {NoStop}%
\bibitem [{\citenamefont {Wright}\ \emph {et~al.}(2013)\citenamefont {Wright},
  \citenamefont {Blakestad}, \citenamefont {Lobb}, \citenamefont {Phillips},\
  and\ \citenamefont {Campbell}}]{wright2013}%
  \BibitemOpen
  \bibfield  {author} {\bibinfo {author} {\bibfnamefont {K.~C.}\ \bibnamefont
  {Wright}}, \bibinfo {author} {\bibfnamefont {R.~B.}\ \bibnamefont
  {Blakestad}}, \bibinfo {author} {\bibfnamefont {C.~J.}\ \bibnamefont {Lobb}},
  \bibinfo {author} {\bibfnamefont {W.~D.}\ \bibnamefont {Phillips}}, \ and\
  \bibinfo {author} {\bibfnamefont {G.~K.}\ \bibnamefont {Campbell}},\ }\href
  {\doibase 10.1103/PhysRevLett.110.025302} {\bibfield  {journal} {\bibinfo
  {journal} {Physical Review Letters}\ }\textbf {\bibinfo {volume} {110}},\
  \bibinfo {pages} {025302} (\bibinfo {year} {2013})}\BibitemShut {NoStop}%
\bibitem [{\citenamefont {Bradley}\ \emph {et~al.}(1997)\citenamefont
  {Bradley}, \citenamefont {Sackett},\ and\ \citenamefont
  {Hulet}}]{bradley1997}%
  \BibitemOpen
  \bibfield  {author} {\bibinfo {author} {\bibfnamefont {C.~C.}\ \bibnamefont
  {Bradley}}, \bibinfo {author} {\bibfnamefont {C.~A.}\ \bibnamefont
  {Sackett}}, \ and\ \bibinfo {author} {\bibfnamefont {R.~G.}\ \bibnamefont
  {Hulet}},\ }\href {\doibase 10.1103/PhysRevLett.78.985} {\bibfield  {journal}
  {\bibinfo  {journal} {Physical Review Letters}\ }\textbf {\bibinfo {volume}
  {78}},\ \bibinfo {pages} {985} (\bibinfo {year} {1997})}\BibitemShut
  {NoStop}%
\bibitem [{\citenamefont {Rakonjac}\ \emph {et~al.}(2016)\citenamefont
  {Rakonjac}, \citenamefont {Marchant}, \citenamefont {Billam}, \citenamefont
  {Helm}, \citenamefont {Yu}, \citenamefont {Gardiner},\ and\ \citenamefont
  {Cornish}}]{rakonjac2016}%
  \BibitemOpen
  \bibfield  {author} {\bibinfo {author} {\bibfnamefont {A.}~\bibnamefont
  {Rakonjac}}, \bibinfo {author} {\bibfnamefont {A.~L.}\ \bibnamefont
  {Marchant}}, \bibinfo {author} {\bibfnamefont {T.~P.}\ \bibnamefont
  {Billam}}, \bibinfo {author} {\bibfnamefont {J.~L.}\ \bibnamefont {Helm}},
  \bibinfo {author} {\bibfnamefont {M.~M.~H.}\ \bibnamefont {Yu}}, \bibinfo
  {author} {\bibfnamefont {S.~A.}\ \bibnamefont {Gardiner}}, \ and\ \bibinfo
  {author} {\bibfnamefont {S.~L.}\ \bibnamefont {Cornish}},\ }\href {\doibase
  10.1103/PhysRevA.93.013607} {\bibfield  {journal} {\bibinfo  {journal}
  {Physical Review A}\ }\textbf {\bibinfo {volume} {93}},\ \bibinfo {pages}
  {013607} (\bibinfo {year} {2016})}\BibitemShut {NoStop}%
\bibitem [{\citenamefont {Sachkou}\ \emph {et~al.}(2019)\citenamefont
  {Sachkou}, \citenamefont {Baker}, \citenamefont {Harris}, \citenamefont
  {Stockdale}, \citenamefont {Forstner}, \citenamefont {Reeves}, \citenamefont
  {He}, \citenamefont {McAuslan}, \citenamefont {Bradley}, \citenamefont
  {Davis},\ and\ \citenamefont {Bowen}}]{sachkou2019}%
  \BibitemOpen
  \bibfield  {author} {\bibinfo {author} {\bibfnamefont {Y.~P.}\ \bibnamefont
  {Sachkou}}, \bibinfo {author} {\bibfnamefont {C.~G.}\ \bibnamefont {Baker}},
  \bibinfo {author} {\bibfnamefont {G.~I.}\ \bibnamefont {Harris}}, \bibinfo
  {author} {\bibfnamefont {O.~R.}\ \bibnamefont {Stockdale}}, \bibinfo {author}
  {\bibfnamefont {S.}~\bibnamefont {Forstner}}, \bibinfo {author}
  {\bibfnamefont {M.~T.}\ \bibnamefont {Reeves}}, \bibinfo {author}
  {\bibfnamefont {X.}~\bibnamefont {He}}, \bibinfo {author} {\bibfnamefont
  {D.~L.}\ \bibnamefont {McAuslan}}, \bibinfo {author} {\bibfnamefont {A.~S.}\
  \bibnamefont {Bradley}}, \bibinfo {author} {\bibfnamefont {M.~J.}\
  \bibnamefont {Davis}}, \ and\ \bibinfo {author} {\bibfnamefont {W.~P.}\
  \bibnamefont {Bowen}},\ }\href {\doibase 10.1126/science.aaw9229} {\bibfield
  {journal} {\bibinfo  {journal} {Science}\ }\textbf {\bibinfo {volume}
  {366}},\ \bibinfo {pages} {1480} (\bibinfo {year} {2019})}\BibitemShut
  {NoStop}%
\bibitem [{\citenamefont {Ellis}\ and\ \citenamefont {Luo}(1989)}]{ellis1989}%
  \BibitemOpen
  \bibfield  {author} {\bibinfo {author} {\bibfnamefont {F.~M.}\ \bibnamefont
  {Ellis}}\ and\ \bibinfo {author} {\bibfnamefont {H.}~\bibnamefont {Luo}},\
  }\href {\doibase 10.1103/PhysRevB.39.2703} {\bibfield  {journal} {\bibinfo
  {journal} {Physical Review B}\ }\textbf {\bibinfo {volume} {39}},\ \bibinfo
  {pages} {2703} (\bibinfo {year} {1989})}\BibitemShut {NoStop}%
\bibitem [{\citenamefont {B{\"u}hler}(2002)}]{buhler2002}%
  \BibitemOpen
  \bibfield  {author} {\bibinfo {author} {\bibfnamefont {O.}~\bibnamefont
  {B{\"u}hler}},\ }\href {\doibase 10.1063/1.1483305} {\bibfield  {journal}
  {\bibinfo  {journal} {Physics of Fluids}\ }\textbf {\bibinfo {volume} {14}},\
  \bibinfo {pages} {2139} (\bibinfo {year} {2002})}\BibitemShut {NoStop}%
\end{thebibliography}%
\end{document}


\title{
Supplemental Material:\\ Universal expansion of vortex clusters in a dissipative two-dimensional superfluid}

\author{Oliver R. Stockdale} \affiliation{ARC Centre of Excellence in Future Low-Energy Electronics Technologies, School of Mathematics and Physics, University of Queensland, St Lucia, QLD 4072, Australia}

\author{Matthew T. Reeves}\affiliation{ARC Centre of Excellence in Future Low-Energy Electronics Technologies, School of Mathematics and Physics, University of Queensland, St Lucia, QLD 4072, Australia}

\author{Xiaoquan Yu}
\affiliation{Graduate School of China Academy of Engineering Physics, Beijing 100193, China}
\affiliation{The Dodd-Walls Centre for Photonic and Quantum Technologies, Department of Physics, University of Otago, Dunedin 9016, New Zealand}

\author{Guillaume Gauthier}\affiliation{ARC Centre of Excellence for Engineered Quantum Systems, School of Mathematics and Physics, University of Queensland, St Lucia, QLD 4072, Australia}

\author{Kwan Goddard-Lee}\affiliation{ARC Centre of Excellence for Engineered Quantum Systems, School of Mathematics and Physics, University of Queensland, St Lucia, QLD 4072, Australia}

\author{Warwick P. Bowen}\affiliation{ARC Centre of Excellence for Engineered Quantum Systems, School of Mathematics and Physics, University of Queensland, St Lucia, QLD 4072, Australia}

\author{Tyler W. Neely}\affiliation{ARC Centre of Excellence for Engineered Quantum Systems, School of Mathematics and Physics, University of Queensland, St Lucia, QLD 4072, Australia}

\author{Matthew J. Davis}
\affiliation{ARC Centre of Excellence in Future Low-Energy Electronics Technologies, School of Mathematics and Physics, University of Queensland, St Lucia, QLD 4072, Australia}
\affiliation{ARC Centre of Excellence for Engineered Quantum Systems, School of Mathematics and Physics, University of Queensland, St Lucia, QLD 4072, Australia}

\maketitle
\section*{Dissipative Vortex Fluid Theory}
\subsection*{Vortex velocity field}
In the universal expanding regime, where $\nabla \rho=0$, the vortex velocity fluid $\vb{v}$ is not needed in order to determine the dynamics of the vorticity distribution. However, for our work to be complete and  self-contained, here we provide the full set of equations that describe a dissipative chiral vortex fluid.    

Dissipation effects due to thermal friction can be taken into account by considering a dissipative point-vortex model of the form
\begin{equation}
    \dot{\vb{r}}_i = \vb{v}_{i}  - \kappa_i  \gamma\, (\vu{z}\times  \vb{v}_{i}), 
    \label{eqn:dampedPVM}
\end{equation}
where the dimensionless mutual friction coefficient $\gamma$ (typically $\ll1 $) characterises the strength of the dissipation. A general dissipative binary vortex fluid theory (containing vortices and anti-vortices) has been formulated in Ref.~\cite{yu2017}. Here we only consider the chiral limit, where $\kappa_i=1$.  We define the vortex density $\rho \equiv \sum_{i}\delta(\vb{r}-\vb{r}_i)$. The conservation of vortex number implies the following continuity equation~\cite{yu2017} 
\bea{\label{dissipativeceq}\partial_t \rho+\nabla \cdot \mathbf{J}'_n =0,}
where the number current $\mathbf{J}'_n=\mathbf{J}_n-\gamma \mathbf{\hat{z}} \times \mathbf{J}_n$ and $\mathbf{J}_n=\sum_i\delta(\vb{r}-\vb{r}_i)\vb{v}_i$.  The vortex velocity field $\vb{v}$ is defined according to the hydrodynamic relation $\mathbf{J}_n\equiv\rho \vb{v}$. 
Re-writing Eq.~\eqref{dissipativeceq}, we obtain the dissipative Helmholtz equation
\bea{\label{dampHwvChiral}{\cal D}^{\hat{v}}_t\rho=-\gamma\left(8\pi \eta \rho^2+\eta \nabla^2 \rho-\mathbf{v} \times \nabla \rho\right),}
where $\hat{\vb{v}}\equiv \mathbf{v}-\gamma\eta \nabla \ln \rho$ and $\eta\equiv \Gamma/(8\pi)=h/(8 \pi m)$.
Following a similar coarse-graining procedure used in Ref.~\cite{yu2017}, we obtain,  in Cartesian components,
\bea{\label{dampeduniversalformchiral}& {\cal D}^{v}_t v_a-\rho^{-1}\partial_b \hat{\Pi}_{ab}+ \partial_a \hat{p}=-8\eta \gamma \pi \rho v_a +\rho^{-1}F_a,}
that governs the dynamics of the vortex velocity field,  where the modified Cauchy stress tensor  $\hat{\Pi}_{ab}=\Pi_{ab}+\Pi'_{ab}$ with
\bea{\Pi_{ab}&=-\rho \tau_{ab}-8\eta^2 \pi \rho^2 \delta_{ab}- \eta^2 \rho \partial_b(\rho^{-1}\partial_a \rho), \\
	\Pi'_{ab}&=\eta\gamma(2\delta_{ab}-1)\Gamma_{aa'}\Gamma_{bb'} \left[\partial_{a'}(\rho v_{b'})+\partial_{b'}(\rho v_{a'})\right],\\
	\tau_{xy}&=\tau_{yx}=\eta(\partial_x v_x-\partial_y v_y),\\
\tau_{xx}&=-\tau_{yy}=-\eta(\partial_x v_y+\partial_y v_x),}
and 
\bea{
F_a&=-\eta \gamma \rho^{-1}\epsilon_{ab} \epsilon_{cd}\partial_b \rho \partial_c (\rho v_d);\\
\hat{p}&=p-2\eta \gamma (g+\bar{g}).} 
Here $\Gamma_{aa}=0$, $\Gamma_{a\neq b}=1$, $\epsilon_{ab}$ is the anti-symmetric tensor ($\epsilon_{xy}=-\epsilon_{yx}=1$), and $g$ is determined by $\partial_{\bar{z}} g=\pi \rho \bar{v}$.   The coefficient $\eta \gamma$ might be identified as the viscosity of the vortex fluid. Since $(1/2){\rm tr}( \Pi'_{ab})=\eta \gamma \nabla \cdot (\rho \mathbf{v})$, it may also play the role of {\em second viscosity}.

The  pressure $\hat{p}$ can be eliminated by taking the curl of Eq.~\eqref{dampeduniversalformchiral} 
\begin{widetext}
\bea{\label{dampeduniversalform}& \epsilon_{ac}\partial_t \partial_c v_a+ \epsilon_{ac} \partial_c(v_b \partial_b v_a)-\epsilon_{ac} \partial_c(\rho^{-1}\partial_b \hat{\Pi}_{ab})
=-8\eta \gamma \pi \epsilon_{ac} \partial_c(\rho v_a) +\epsilon_{ac} \partial_c(\rho^{-1}F_a).}
Hence the complete set of equations for dissipative chiral vortex fluids reads 
\bea{\label{dampHwvChiral-2}\partial_t \rho+ \left(\mathbf{v}-\gamma\eta \nabla \ln \rho \right) \cdot \nabla \rho&=-\gamma\left(8\pi \eta \rho^2+\eta \nabla^2 \rho-\mathbf{v} \times \nabla \rho\right). \\
\epsilon_{ac}\partial_t \partial_c v_a+ \epsilon_{ac} \partial_c(v_b \partial_b v_a)-\epsilon_{ac} \partial_c(\rho^{-1}\partial_b \hat{\Pi}_{ab})&=-8\eta \eta^{\ast} \pi \epsilon_{ac} \partial_c(\rho v_a) +\epsilon_{ac} \partial_c(\rho^{-1}F_a),\\
\partial_a v_a&=0.}

\clearpage

\end{widetext}

\subsection*{Dynamics of density perturbations}

We demonstrate that the Rankine vortex solution
\begin{align}
    \rho(t) = \left(\rho_0^{-1} + \gamma\Gamma t\right)^{-1}\label{Rankine}
\end{align}
is an asymptotic solution of Eq.~(3) of the main text (anomalous hydrodynamical equation for the vortex density) in the long time limit, independent of initial vortex distributions. Consider a local density perturbation $\rho~=~\rho_{a}~+~\delta \rho $ to the universal expansion solution, where $\rho_a$ is given by \eq{Rankine}, $|\delta \rho| \ll |\rho_a|$, and $\delta \rho (|\vb{r}|\rightarrow \infty)=0$. For simplicity, we assume that $\delta \rho$ has cylindrical symmetry, namely, $\delta \rho(\mathbf{r})=\delta\rho(r)$. Keeping the leading order terms in $\delta \rho$, we obtain 
\begin{align}
\frac{\partial \delta \rho}{\partial t}  +\frac{\gamma \Gamma}{8\pi}\partial^2_r\delta \rho +\gamma\left(2  \Gamma \rho_a \delta \rho+ v_a \partial_r \delta \rho \right) =0,
\label{perturbation}
\end{align}
where $v_a=r \Gamma \rho_a/2 $.
The last two terms in Eq.~\eqref{perturbation} become less relevant at long times as the coefficients in front of $\delta \rho$ and $\partial_r \delta \rho$ are proportional to $\rho_a$, which tends to zero in the long time limit. Hence, the evolution of the vorticity distribution will eventually be dominated by a diffusion equation with a negative diffusion coefficient. The only physical solution (i.e.\ no singularities during the time evolution) is therefore a constant density. Indeed, solving \eq{perturbation} numerically with several different initial vortex density distributions shows that $\delta\rho(t\rightarrow\infty)\rightarrow0$.

\begin{figure}
    \centering
\includegraphics[width=\columnwidth]{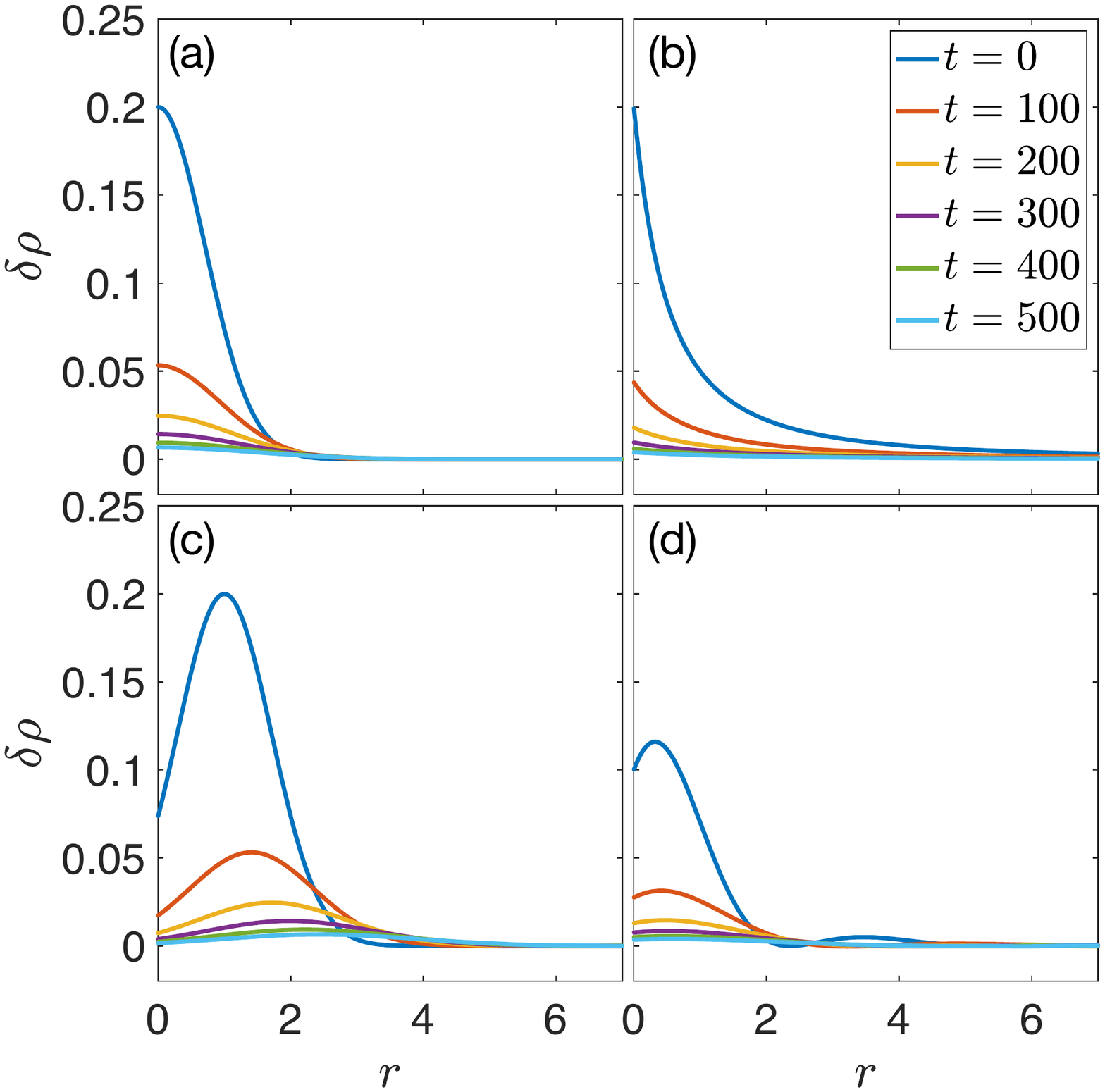}
    \caption{Time evolution of perturbation $\delta \rho$ for different initial distributions. (a) $\delta \rho (r,0)=0.2 \exp(-r^2)$; (b) $\delta \rho (r,0)=0.2/(1 + r)^2$; (c) $\delta \rho (r,0)=0.2 \exp(-(r - 1)^2)$; (d) $\delta \rho (r,0)=0.1 (\sin 2 r + 1)\exp(-r)$. Here we choose $\Gamma=1$, $\rho_0=1$ and $\gamma=0.01$.}
    \label{fig:perturbations}
\end{figure}

Solving \eq{perturbation} numerically for a variety of initial conditions,  we find that for all cases the perturbation decays over time and reduces to zero across the fluid. Figure~(\ref{fig:perturbations}) shows four
examples where the initial perturbation takes a different form, as specified in the figure caption. Here the boundary conditions are: $\delta \rho$ regular at $r=0$ and $\delta \rho (r\rightarrow \infty)\rightarrow 0$.

This result suggests that the top-hat distribution is an attractor of the dissipative vortex fluid dynamics, and hence further supports our conclusion that any initial distribution of vortex density will eventually tend towards a Rankine vortex.\\

\section*{Vortex distributions and crystallization}

\subsection*{Emerging Abrikosov lattice}
\begin{figure*}
    \centering
    \includegraphics[width=15cm]{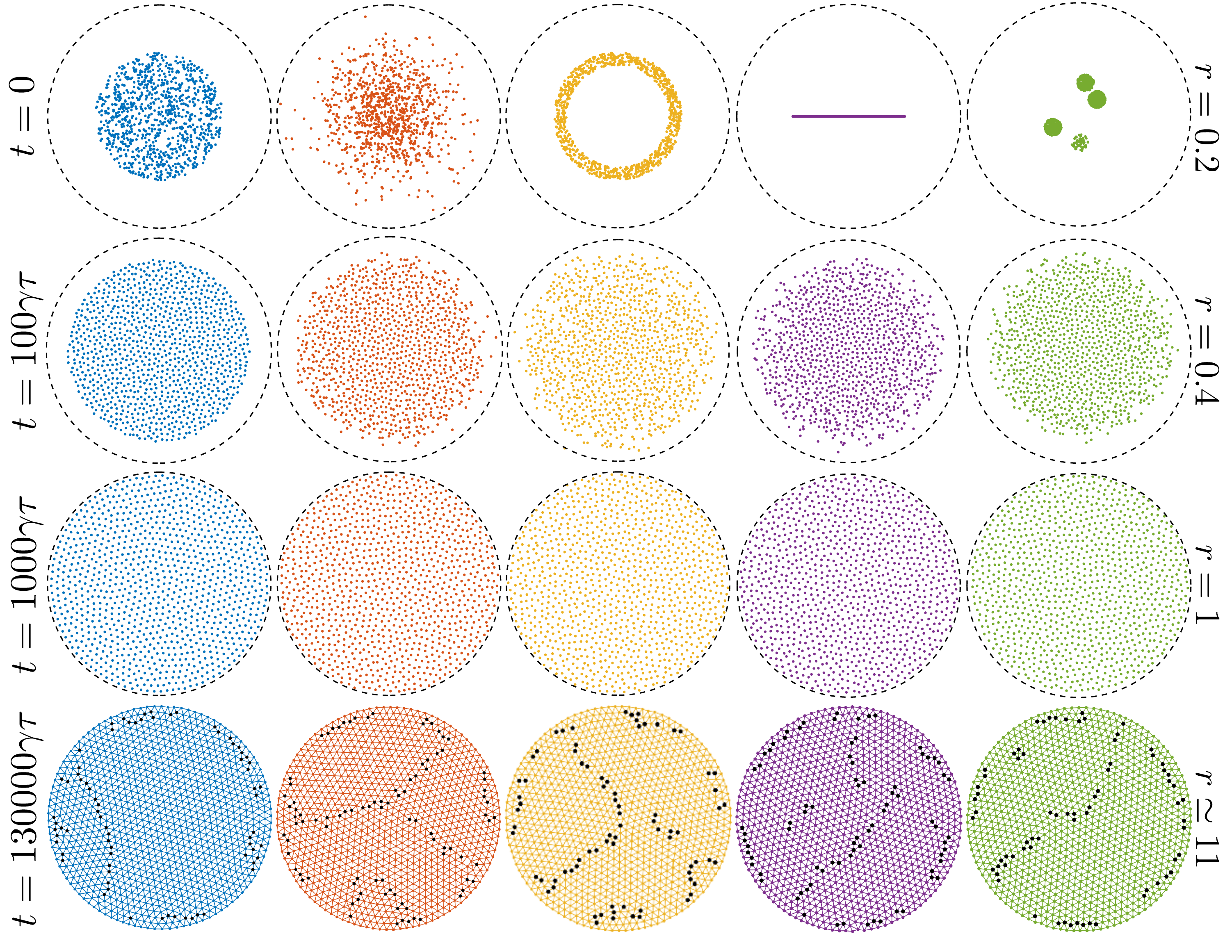}
    \caption{Exemplar vortex distributions at different times during the point vortex model simulation of the cluster expansion. We show the top-hat, Gaussian, and ring distributions in columns one, two and three respectively as presented in the main text. We additionally show two non-axisymmetric cases of a cluster expansion in the fourth and fifth columns.  The simulation time is indicated on the left, and the radius of the dashed lined around the cluster is indicated on the right. In the final row, at $t=130000\gamma\tau$, we have drawn lines between the nearest neighbour to demonstrate the crystallization of the cluster. Lattice dislocations, where vortices have either five or seven nearest neighbours, are indicated by black dots.}
    \label{all}
\vspace*{\floatsep}
    \centering
    \includegraphics[width=15cm]{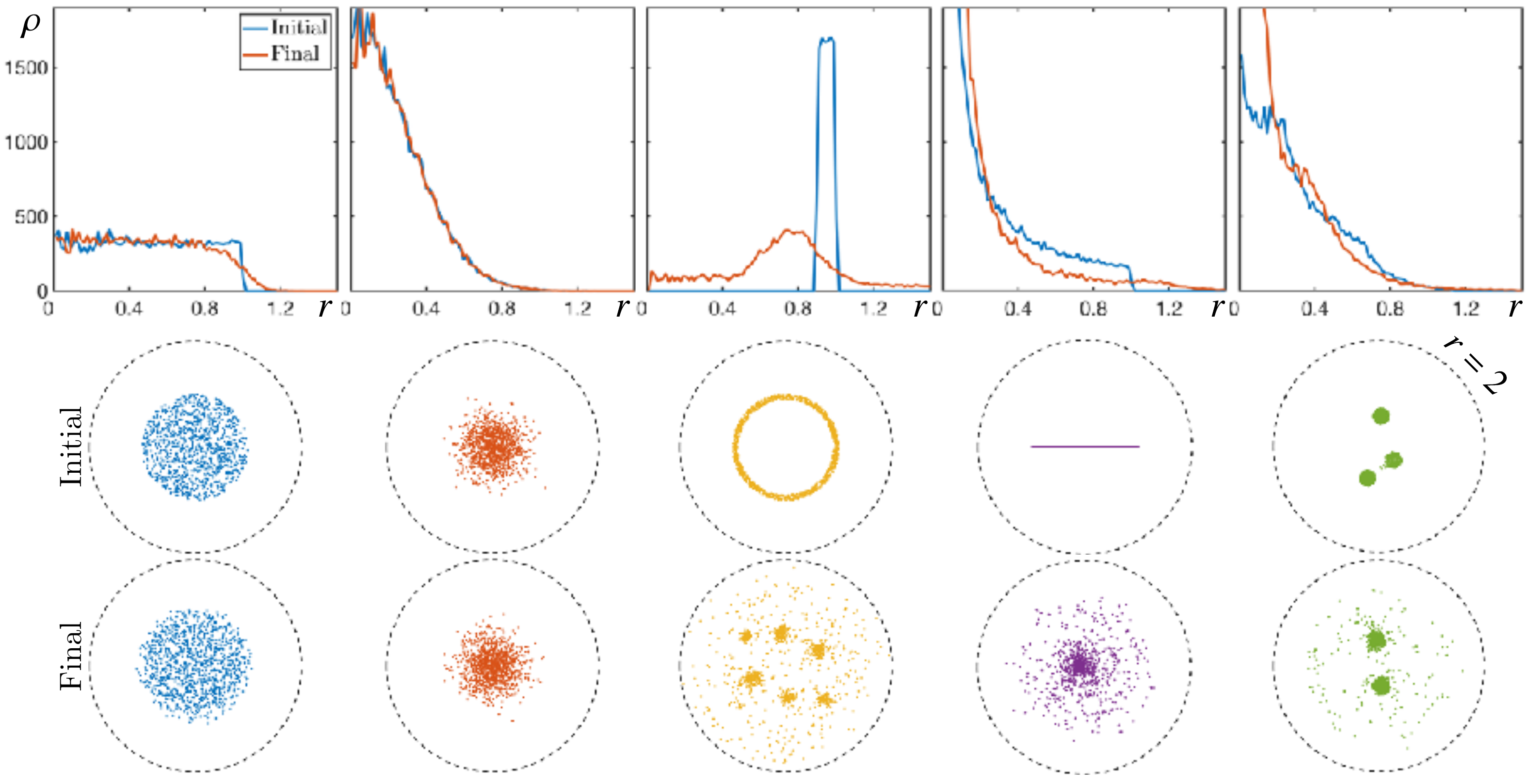}
    \caption{Evolution of vortex clusters without dissipation (i.e., $\gamma=0$). Each column is for a  different distribution, corresponding to those of \fig{all}. The upper row is the radial density, and the middle and bottom rows are exemplar snapshots of the cluster at the beginning and end of the evolution (as labelled). The initial clusters are evolved for $t=10\tau$, the same amount of time as for the final states in Fig.~1(c) of main text.} 
    \label{zeroG}
\end{figure*}

In \fig{all} we display examples of the vortex distributions from simulations of the point vortex model. The first three columns correspond to the three examples presented in the main text.  The last two columns present results for additional initial distributions without axial symmetry. Each row represents a time, indicated on the left, and the radius of the dashed lined around the cluster is indicated on the right. In the bottom row we highlight the crystallization of the cluster by connecting nearest neighbour vortices. Vortices that have either five or seven nearest neighbours are coloured black to identify the dislocations in the crystal. In future work, we will 
characterize the emergence of these dislocations.

\subsection*{Zero dissipation}
To emphasise the dramatic difference between the dynamics of large clusters of vortices in the presence or absence of dissipation,
we plot the evolution of the five different initial distributions when the dissipation is zero (i.e., solve \eq{eqn:dampedPVM} for $\gamma=0$). In \fig{zeroG}, it is 
clear that the Rankine vortex does not emerge in these situations, and instead
the final vortex distributions are strongly dependent
upon the initial conditions.

\subsection*{Vortex-vortex correlation function}
\begin{figure}
    \centering
    \includegraphics[width=\columnwidth]{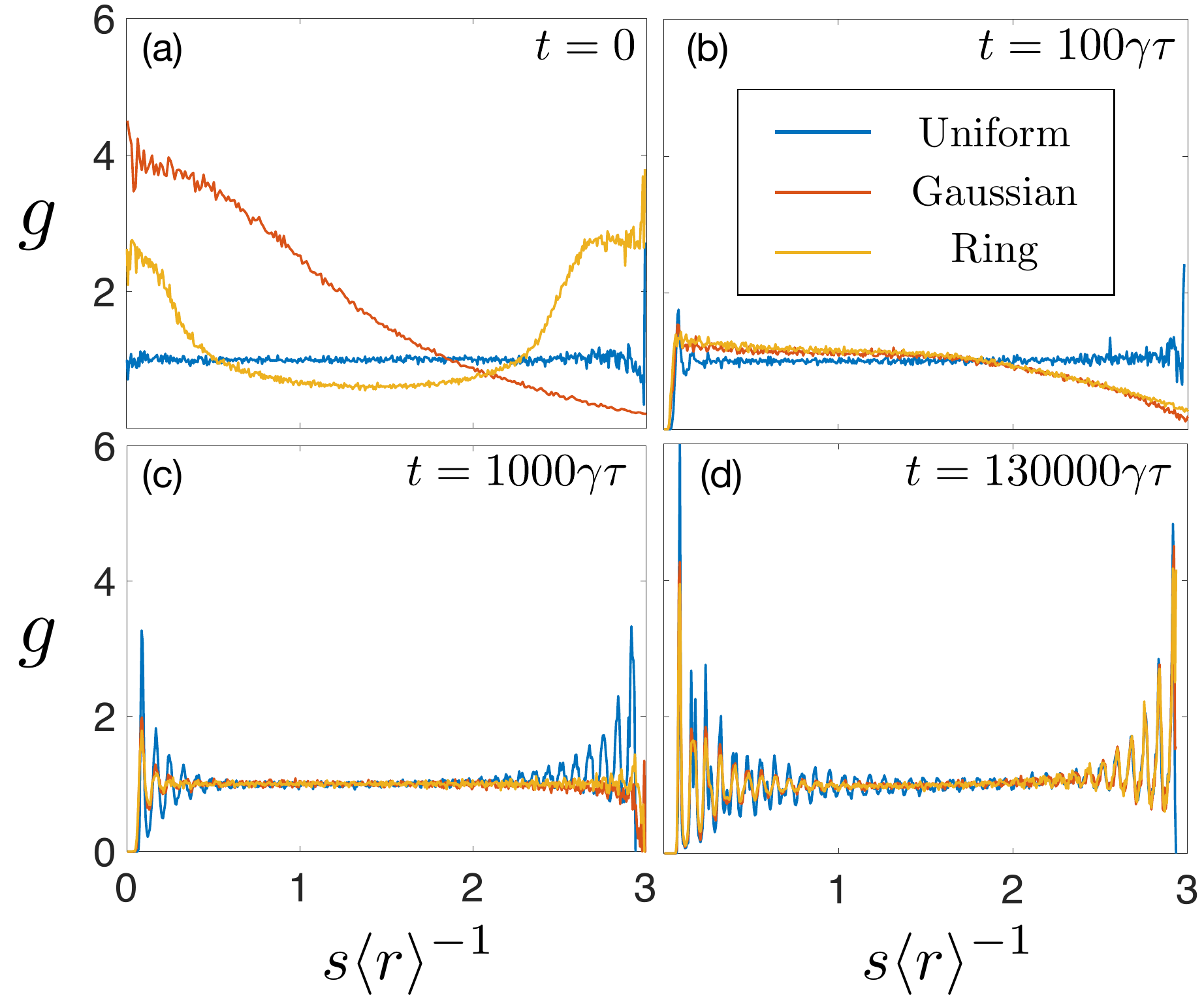}
    \caption{Vortex-vortex correlation function of the cluster for times corresponding to those in \fig{all} for the initial top-hat (uniform), Gaussian, and ring distributions.}
    \label{fig:correlation}
\end{figure}

To further quantify the ordering and crystallization of the vortex lattice, we calculate the vortex-vortex correlation function
\begin{align}
    g(s) = \frac{1}{\rho \mathcal{N}}\sum_{i}\sum_{j\neq i}\frac{1}{s(2\pi-\Delta\theta_i)}\delta(s-r_{ij}),
\end{align}
where $s$ is the spatial separation of a pair of vortices, $\mathcal{N}$ is the number of vortices contributing to  separation $s$, and $\Delta\theta_i$ is the accessible angle at distance $s$ for a given vortex assuming a circular distribution~\cite{skaugen2017}.  No correlation is indicated by $g(s)=1$, as is the case for the initial top-hat distribution [blue curve in \fig{fig:correlation}(i)]. However, periodic correlations emerge for all ensembles during expansion as seen in \fig{fig:correlation}(iii) at $t=1000\gamma\tau$. In the long time limit the separation of vortices on opposite sides of the cluster also become correlated. The correlation functions for the three ensembles exhibit the same correlation period, but differ in the strength of the correlation for the same evolution time.

\section*{Experimental procedure and further results}

The experimental system has been previously described elsewhere in Refs.~\cite{gauthier2016,gauthier2018}. We trap a $N\sim 2.2 \times 10^6$  $^{87}$Rb Bose-Einstein, with a condensate fraction of $\sim80\%$, in a gravity-compensated optical potential. The horizontal confinement of the atoms in the $x$-$y$ plane is provided by a repulsive blue-detuned optical dipole potential that results from the direct imaging of a digital micro-mirror device; here it is configured to produce a disc-shaped trap with a $50~\mu$m radius. This results in the approximately hard-walled confinement of the BEC with a near-uniform density.  In the vertical direction a red-detuned optical dipole potential provides harmonic trapping with frequency $\omega_z = 2\pi \times 108$ Hz, leading to a  vertical Thomas-Fermi radius of $6~\mu$m.  The healing  length of the BEC is  $\xi \sim 500$~nm ~\citep{gauthier2018}.

\begin{figure*}[t]
    \centering
    \includegraphics[width=17cm]{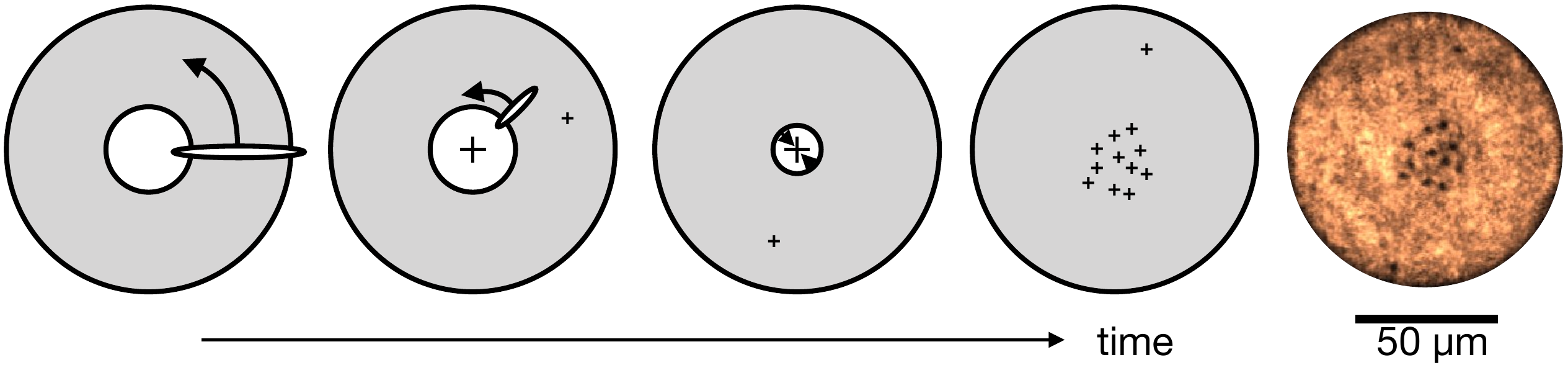}
    \caption{Schematic of the procedure used to create the initial vortex cluster. An elliptical stirrer is swept around the annulus as its length is shortened over time, creating same-signed vortices (denoted by a $+$) which become pinned to the central barrier.  However, some vortices can become stray and remain in the bulk of the fluid. The radius of the central barrier is then shrunk to zero, and the pinned circulation (denoted by large $+$ at the center) splits into singly charged vortices which are free to expand. (Right) Experimental image after a 2.75~s hold time and 3~ms time-of-flight expansion, showing resolved vortex cores as dark holes in the atom density.}
    \label{schem}
\end{figure*}

\subsection*{Initial state preparation}
A schematic of the preparation of the initial vortex cluster  shown in \fig{schem}. We form the BEC in the disc trap, before transferring it to an annular trap by ramping-on an additional repulsive central barrier with a radius of $R_0 = 15~\mu$m over 200~ms. Simultaneously, an elliptical stirring barrier with a major and minor axis of $50~\mu$m and $2~\mu$m, respectively, that crosses the annulus is introduced, resulting in a split ring. Following previous techniques for stirring persistent currents in ring-trapped BECs~\cite{eckel2014,wright2013}, the stirring barrier is linearly accelerated at $~980~\mu\textrm{m}\,\textrm{s}^{-2}$ for a time of 400~ms around the annulus. While still accelerating at the same rate, the barrier height is then linearly to zero by reducing both the barrier width and length over 100~ms, effectively removing the stirring barrier through to the central barrier. After a 400~ms period of equilibration in the annular trap, the central barrier is removed over 200~ms by linearly reducing its radius to zero. This results in a high-energy cluster of $\sim11$ vortices near the trap center. During the removal of the stirrer, sometimes one or two  ``stray'' vortices of the same sign are produced away from the main cluster, as described in the main text.

\subsection*{BEC imaging and vortex detection} 
High-resolution images of the BEC and vortex cores are obtained as in Ref.~\cite{gauthier2018}, where a short 3~ms time-of-flight allows the vortex cores to expand and become visible using darkground Faraday imaging~\cite{bradley1997}. The radial distribution of the condensate is essentially unchanged from this expansion. An example image is shown in Fig.~\ref{schem}. The vortex positions were identified using a Gaussian fitting algorithm~\citep{rakonjac2016, gauthier2018}. 

\begin{figure}[t]
    \centering
    \includegraphics[width=\columnwidth]{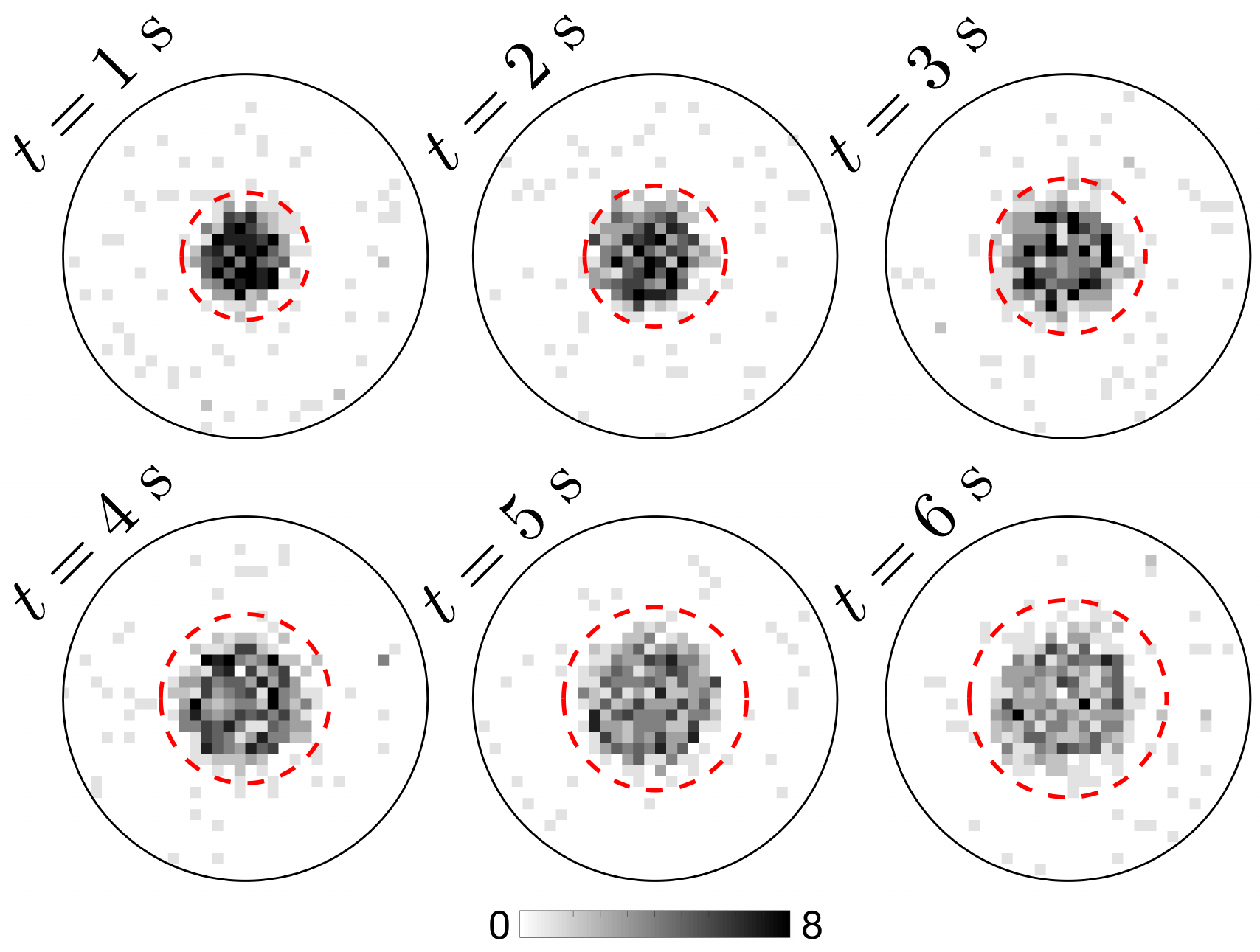}
    \caption{Vortex position histograms for the cluster expansion experiment for $\sim40$ sets of data at times indicated in the upper left corner.  The red dashed line indicates the cut-off radius for the cluster analysis (see text for details).}
    \label{expt}
\end{figure}

\subsection*{Experimental histograms}
In \fig{expt} we show histograms of the vortex positions in the experiment at a single time, as compared to the main text where we have averaged the vortex distributions over one second. As in the main text, we plot all vortices that have been detected, and show the cut-off radius beyond which the vortices are identified as not being part of the cluster for subsequent analysis. However, when running point vortex simulations to calculate the evolution of the lattice disorder $\sigma_g$ in the main text we include these stray vortices. We find the results for $\sigma_g$ presented in the main text are insensitive to small changes in the cut-off radius.

\begin{figure}[t]
    \centering
    \includegraphics[width=\columnwidth]{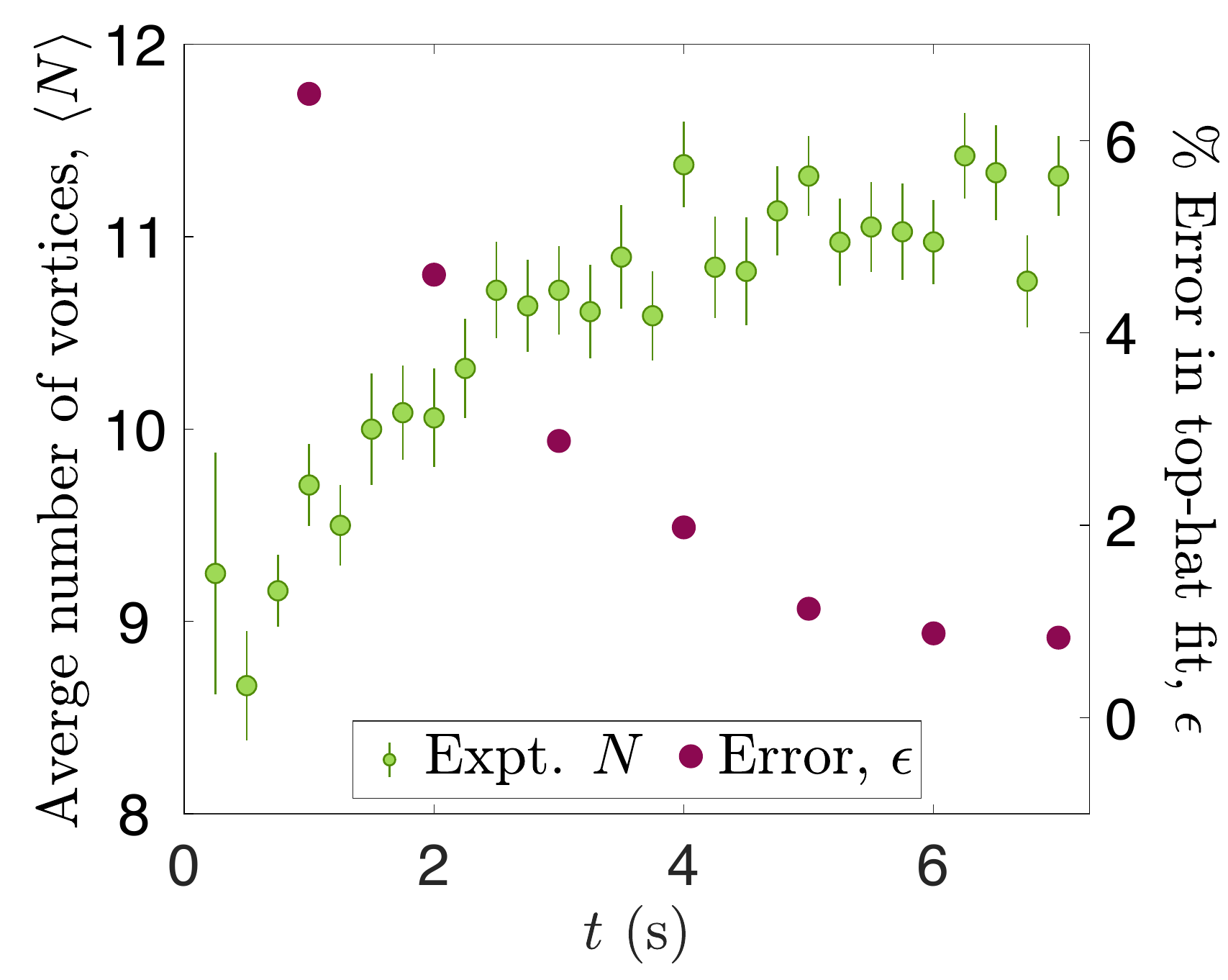}
    \caption{Left axis: Average number of vortices counted as a function of time (green points with error bars calculated from standard error). This increases initially due to the difficulty in resolving individual vortex cores at early times as described in the text. Right axis: Percentage error between the top-hat fit and the radial density histograms as shown in Fig.~3 of the main text.}
    \label{fig:extraexpt}
\end{figure}

\subsection*{Vortex number and top-hat fit} 
In \fig{fig:extraexpt} we plot the average number of vortices counted at each time the condensate is imaged. The number of vortices initially increases due to the difficulty in resolving the individual positions when the vortices are tightly clustered at the center of the disk. On the right axis of \fig{fig:extraexpt} we plot the error between the top-hat fit and the one-dimensional radial density histograms shown in the main text. The radius of these top-hat fits are calculated from the measured average radius shown in Fig.~3(g) in the main text. The error is calculated as
\begin{align}
    \epsilon=\frac{1}{n}\frac{|\text{Expt}-\text{Top-hat}|}{\text{Expt}},
\end{align}
where $n$ is the number of bins used in the histogram. 
We can see that the error decreases as the cluster expands, supporting the conclusion that the cluster is approaching the universal expanding regime.

\section*{Point vortex simulations for finite-sized systems}

The vortex cluster expansion experiments we describe in the main text are performed with $\sim11$ vortices in an atomic gas Bose-Einstein condensate (BEC).  The superfluid is confined in a flat-bottomed, circular trap of radius $R$ using the experimental techniques described in Ref.~\cite{gauthier2016}.  Similar systems have been realised for superfluid helium~\cite{sachkou2019,ellis1989}.

The vortex fluid hydrodynamic theory, however, is for an infinite system with no boundaries in which the vortex density is coarse-grained and treated as a continuous quantity. Thus an important question is --- how applicable is the vortex fluid theory to the experimental observations?  In this section we compare simulations of point vortex dynamics in the circular trap with the predictions of the time-dependence of the cluster radius and energy from the vortex fluid theory.

The Hamiltonian for $N$ vortices at positions $\mathbf{r}_i$ in a circular domain of radius $R$ is~\cite{buhler2002}
\begin{align}
H = &- \frac{1}{4\pi}\sum_{i<j}^{N,N}\Gamma_i\Gamma_j\ln\qty(\frac{r_{ij}^2}{R^2}) + \frac{1}{4\pi}\sum_{i = 1}^N\Gamma_i^2\ln\qty(\frac{R^2-r_i^2}{R^2})\nonumber  \\ &+ \frac{1}{4\pi}\sum_{i<j}^{N,N}\Gamma_i\Gamma_j\ln\qty(\frac{R^4 - 2R^2\vb{r}_i\cdot\vb{r}_j+r_i^2r_j^2}{R^4}),\label{hamilCirc}
\end{align}
where $r_i = |\vb{r}_i|$ is the magnitude of the radial position of the vortex, and $r_{ij} = |\vb{r}_i - \vb{r}_j|$. The first term of \eq{hamilCirc} describes the energy due to vortex-vortex interactions.  The second and third terms are from the interactions with fictitious image vortices enforce the boundary condition $\mathbf{u} \cdot \hat{\mathbf{n}} |_{r=R} = 0$, which ensures the superfluid flow normal to the boundary is zero.

Hamilton's equations of motion for this system are
\begin{align}
    \Gamma_i\dv{x_i}{t} &= \frac{\partial H}{\partial y_i}, & \Gamma_i\dv{y_i}{t} &= -\frac{\partial H}{\partial x_i}, \label{HamiltonsEquations}
\end{align}
which yield the equation of motion for vortices in a circular disk
\begin{align}
\mathbf{v}_{i} = \frac{1}{2\pi}
\sum_{j\neq i }\frac{\Gamma_j}{r_{ij}^2}
\begin{pmatrix}
-y_{ij}\\
x_{ij}\\
\end{pmatrix} 
+ \frac{1}{2\pi} \sum_j\frac{\bar{\Gamma}_j}{\bar{r}_{ij}^2}
\begin{pmatrix}
-\bar{y}_{ij}\\
\bar{x}_{ij}\\
\end{pmatrix},\label{eqnOfMotionCirc}
\end{align}
where $x_{ij} = x_i-x_j$, $y_{ij} = y_i-y_j$, and $r_{ij}^2 = x_{ij}^2 + y_{ij}^2$. The first term of \eq{eqnOfMotionCirc} arises from the flow field of the other vortices, whilst the second  term arises from the image vortices. The barred terms $\bar{x}_{ij} = x_i-\bar{x}_j$, $\bar{r}^2_{ij} = \bar{x}_{ij}^2+\bar{y}_{ij}^2$, etc., correspond to the positions of the image vortices, with circulation $\bar \Gamma_j = -\Gamma_j$, located outside the disk at the inverse point $\vb{\bar{r}}_i = {R^2\vb{r}_i}/{|\vb{r}_i|^2}$. Using \eq{eqnOfMotionCirc} for the bounded vortex velocities in Eq.~(2) of the main text, we solve for the dynamics of bounded vortices in the dissipative regime.  

\begin{figure}
    \centering
    \includegraphics[width=\columnwidth]{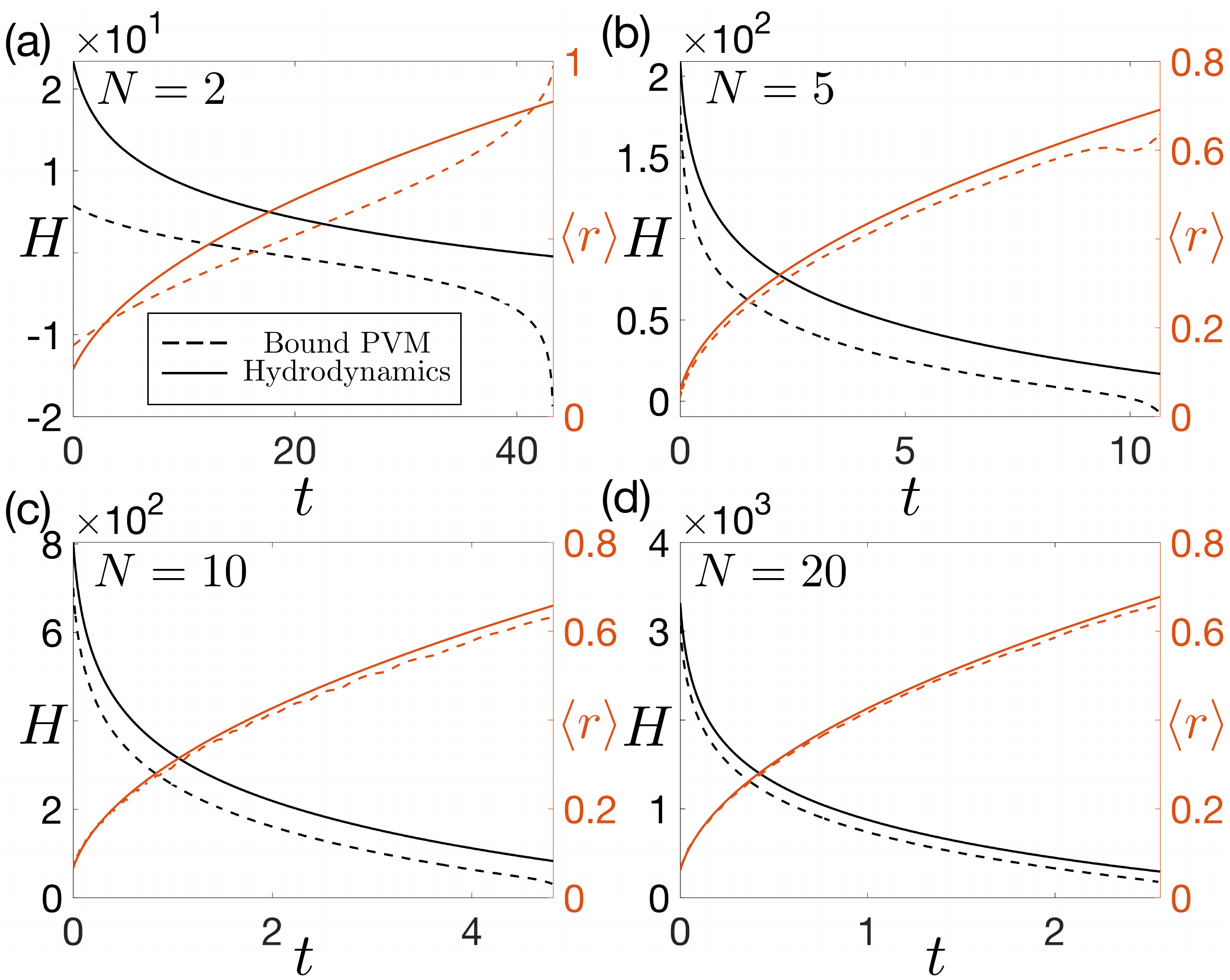}
    \caption{Comparison of radius and energy of point vortex cluster expansion with the predictions of the vortex fluid hydrodynamics.  (a) A system of $N=2$ vortices initially equidistant from origin separated by $\theta=\pi$. The black curves correspond to the left $y$-axis and are the energy of the vortex cluster, whilst the orange curves correspond to the right $y$-axis and indicate the average radius of the vortex cluster. Dashed lines are from simulations of the finite-size point vortex model, and the solid lines are the solution to the vortex fluid hydrodynamics. (b) $N=5$ vortices. (c) $N=10$ vortices. (d) $N=20$ vortices. 
    }
    \label{fig:energy}
\end{figure}

When a vortex approaches the boundary and pairs with its image, the energy of the system reduces, whilst the velocity of the vortex increases dramatically. This can be seen in \fig{fig:energy}(a), where we plot the energy (left axis) and radius (right axis) of two positive vortices in the bounded point vortex model. Unsurprisingly, the hydrodynamic approximation is a poor description of the radius of the vortex pair (plotted as the solid curve) as the core assumption is that vortices are densely packed so the vorticity can be coarse grained. Comparing this result with \fig{fig:energy}(b), where we plot the energy and radius of an expanding cluster of $N=5$ vortices~ (the initial condition is drawn from a uniform distribution, and sits within $r=0.1R$), we see that the boundary has much less of an effect upon the dynamics. As there are more vortices within the center of the disk, the energy is dominated by vortex-vortex interactions, unlike the $N=2$ case where the images strongly influence the dynamics. As image vortices are positioned at the inverse point, their effect upon the system reduces significantly  for vortices far from the boundary. As a result, we begin to see signatures of the anomalous hydrodynamics [i.e.\ $\langle r\rangle\propto\sqrt{t}$ and $H\propto-\ln(t)$] despite there only being $N=5$ vortices.

Further increasing the number of vortices to $N=10$ in \fig{fig:energy}(c), the difference between the bound simulations and the free-space anomalous hydrodynamics is further reduced. This particular result is close to the experimental system, and we see good agreement between the vortex fluid theory and bound point vortex simulation. For $N=20$ [\fig{fig:energy}(c)], the expansion of the cluster in the point vortex model is close to the vortex fluid prediction. For even larger numbers the difference continues to decrease, and so in conclusion, it seems that for a sufficiently large vortex number ($N>20$), an expanding cluster in a bounded circular domain can be closely approximated as a free space expansion until the cluster radius nears the boundary.

There is a distinct difference, however, between the energy in the point vortex simulation and that of the anomalous hydrodynamic solution. Given the anomalous hydrodynamics assumes $N\gg1$, we find there are finite $N$ effects for small number of vortices. For $N>10$, we find that scaling the point vortex simulation energy (and indeed the experiment, seen in the main text) by $N^2-2N$ gives excellent agreement with the scaled hydrodynamic energy $H/N^2$. It can be seen that the difference vanishes in the $N\rightarrow \infty$ limit.

\bibliography{references.bib}